\documentclass[aps,prd,amsmath,floats,floatfix,twocolumn,
  superscriptaddress,nofootinbib,showpacs]{revtex4-2}

\usepackage{xcolor}
\usepackage{hyperref}
\hypersetup{
  colorlinks,
  citecolor={cyan!50!black},
  linkcolor={orange!60!black},
  urlcolor=violet,
}
\usepackage{orcidlink}
\usepackage{graphicx}
\usepackage{amsmath,amssymb}
\usepackage{mathtools}  %
\usepackage{tensor}
\usepackage{bm} %
\usepackage{commath}  %
\usepackage{physics}
\usepackage{microtype}
\usepackage{enumitem}
\usepackage{csquotes}
\usepackage[binary-units]{siunitx}
\usepackage[capitalise]{cleveref}
\crefname{section}{Sec.}{Sec.}
\Crefname{section}{Section}{Sections}
\newcommand{\ccite}[1]{Ref.~\cite{#1}}
\newcommand{\ccites}[1]{Refs.~\cite{#1}}
\newcommand{\Ccite}[1]{Reference~\cite{#1}}
\newcommand{\Ccites}[1]{References~\cite{#1}}
\newcommand{\tikzinput}[1]{
  \includegraphics{#1.pdf}
}
\usepackage[caption=false]{subfig}

\setlist[description]{font=\normalfont\itshape}

\newcommand{\defeq}{\coloneqq}

\newcommand{\dgF}{\mathcal{F}}
\newcommand{\dgFi}[1]{\tensor{\dgF}{_{\!{#1}}^i}}
\newcommand{\dgFj}[1]{\tensor{\dgF}{_{\!{#1}}^j}}
\newcommand{\dgS}{\mathcal{S}}
\newcommand{\dgb}{b}
\newcommand{\dgf}{f}

\newcommand{\bas}{\psi}
\newcommand{\lagr}{\ell}
\newcommand{\jac}{\mathrm{J}}
\newcommand{\invjac}{{(\jac^{-1})}}
\newcommand{\surf}[1]{{#1}^\Sigma}

\newcommand{\dgA}{\mathcal{A}}
\newcommand{\dgM}{M}
\newcommand{\dgD}{D}
\newcommand{\dgMD}{M\!\!D}
\newcommand{\dgL}{L}
\newcommand{\dgML}{M\!\!L}
\newcommand{\prol}{P}
\newcommand{\restr}{R}

\newcommand{\penaltyparam}{C}

\newcommand{\spectre}{\texttt{SpECTRE}}
\newcommand{\spec}{\texttt{SpEC}}
\usepackage[shortcuts]{extdash}
\newcommand{\vcycle}{V\=/cycle}

\newcommand{\opA}{\dgA}
\newcommand{\opb}{\dgb}
\newcommand{\grd}[1]{\underline{#1}}

\begin{document}

\title{A scalable elliptic solver with task-based parallelism for the \spectre{}
numerical relativity code}

\newcommand{\aei}{\affiliation{Max Planck Institute for Gravitational Physics
(Albert Einstein Institute), Am M{\"u}hlenberg 1, Potsdam 14476, Germany}}
\newcommand{\caltech}{\affiliation{Theoretical Astrophysics, Walter Burke
Institute for Theoretical Physics, California Institute of Technology, Pasadena,
California 91125, USA}}
\newcommand{\cornell}{\affiliation{Cornell Center for Astrophysics and Planetary
Science, Cornell University, Ithaca, New York 14853, USA}}
\newcommand{\fullerton}{\affiliation{Nicholas and Lee Begovich Center for
Gravitational-Wave Physics and Astronomy, California State University,
Fullerton, Fullerton, California 92831, USA}}

\author{Nils L. Vu\,\orcidlink{0000-0002-5767-3949}} \email{owls@nilsvu.de} \aei
\author{Harald P. Pfeiffer\,\orcidlink{0000-0001-9288-519X}} \aei

\author{Gabriel S. Bonilla\,\orcidlink{0000-0003-4502-528X}} \fullerton
\author{Nils Deppe\,\orcidlink{0000-0003-4557-4115}} \caltech
\author{Fran\c{c}ois H\'{e}bert\,\orcidlink{0000-0001-9009-6955}} \caltech
\author{Lawrence E. Kidder\,\orcidlink{0000-0001-5392-7342}} \cornell
\author{Geoffrey Lovelace\,\orcidlink{0000-0002-7084-1070}} \fullerton
\author{Jordan Moxon\,\orcidlink{0000-0001-9891-8677}} \caltech
\author{Mark A. Scheel\,\orcidlink{0000-0001-6656-9134}} \caltech
\author{Saul A. Teukolsky\,\orcidlink{0000-0001-9765-4526}} \caltech \cornell
\author{William Throwe\,\orcidlink{0000-0001-5059-4378}} \cornell
\author{Nikolas A. Wittek\,\orcidlink{0000-0001-8575-5450}} \aei
\author{Tom W\l{}odarczyk\,\orcidlink{0000-0003-0005-348X}} \aei

\date{\today}

\begin{abstract}
Elliptic partial differential equations must be solved numerically for
many problems in numerical relativity, such as initial data for every simulation
of merging black holes and neutron stars.
Existing elliptic solvers can take multiple days to solve these problems at high
resolution and when matter is involved, because they are either hard to
parallelize or require a large amount of computational resources.
Here we present a new solver for linear and nonlinear elliptic problems that is
designed to scale with resolution and to parallelize on
computing clusters. To achieve this we employ a discontinuous Galerkin
discretization, an iterative multigrid-Schwarz preconditioned Newton-Krylov
algorithm, and a task-based parallelism paradigm. To accelerate convergence of the elliptic
solver we have developed novel subdomain-preconditioning techniques.
We find that our multigrid-Schwarz preconditioned elliptic solves achieve
iteration counts that are independent of resolution, and our task-based parallel
programs scale over 200 million degrees of freedom to at least a few thousand
cores. Our new code solves a classic initial data problem for binary black holes
faster than the spectral code \spec{} when distributed to only eight
cores, and in a fraction of the time on more cores.
It is publicly accessible in the next-generation \spectre{}
numerical relativity code.
Our results pave the way for highly parallel elliptic solves in numerical
relativity and beyond.
\end{abstract}

\maketitle

\section{Introduction}

Solving elliptic partial differential equations (PDEs) numerically is important in many
areas of science, including numerical relativity~\cite{BaumgarteShapiro}.
All numerical time evolutions
begin with initial data that capture the physical scenario to be evolved, and
the initial data must typically satisfy a set of constraint equations formulated
as elliptic PDEs. Specifically, to construct initial data for
general-relativistic simulations of black holes and neutron stars we must solve
the Einstein constraint equations, which admit formulations as elliptic
PDEs~\cite{Cook2000lrr, Pfeiffer2004-oo}.
Binary configurations of black holes and neutron stars enjoy particular
prominence as primary sources for gravitational-wave detectors, and numerical
simulations of these systems play an essential role in their
observations~\cite{LIGOScientific:2016aoc, LIGOScientific:2017vwq,
LIGOScientific:2021qlt, LIGOScientific:2018mvr, LIGOScientific:2020ibl,
LIGOScientific:2021djp}.

To construct initial data for general-relativistic simulations, the numerical
relativity (NR) community has put considerable effort towards developing
numerical codes that solve elliptic problems. Most of the existing codes employ
spectral methods to discretize the elliptic equations, such as
\texttt{LORENE}~\cite{Lorene, Grandclement:2006ht} and
\texttt{TwoPunctures}~\cite{Ansorg:2004ds,Ansorg:2005bp},
\texttt{Spells}~\cite{Pfeiffer2003-mt, Ossokine:2015yla, Foucart:2008qt,
Tacik2015-iz, Tacik2016-me} that is part of \spec{}~\cite{spec}, as well as
\texttt{SGRID}~\cite{Dietrich:2015pxa,Tichy:2019ouu},
\texttt{KADATH}~\cite{Grandclement2010-wk,Papenfort:2021hod} and
\texttt{Elliptica}~\cite{Rashti:2021ihv}. The \texttt{COCAL}~\cite{Uryu2012-bo,
Tsokaros:2015fea} code employs finite-difference methods, and
\texttt{NRPyElliptic}~\cite{Assumpcao2021-es} a hyperbolic relaxation
scheme~\cite{Ruter2017-gq}.
All of these codes
vary significantly in the numerical methods employed to solve the discretized
equations. For example, \spec{} uses the \texttt{PETSc} library to perform
an iterative matrix-free solve with a custom preconditioner~\cite{Pfeiffer2003-mt}, whereas
\texttt{KADATH} and \texttt{Elliptica} construct explicit matrix
representations and invert the matrices directly~\cite{Grandclement2010-wk,Rashti:2021ihv}.

While successful in constructing initial data for many general-relativistic
scenarios, these codes can still take a significant amount of time or require
excessive computational resources to solve the elliptic problems. For example,
the \spec{} and \texttt{SGRID} codes typically require a few hours to days
to solve for initial data that involves orbiting neutron stars, at a resolution
required for state-of-the-art simulations, using $\mathcal{O}(10)$ cores for the
computation. \texttt{KADATH}, on the other
hand, quote a few hours to solve for low-resolution initial data involving
orbiting neutron stars on about 128 cores, and \enquote{a larger timescale} and
more cores for higher resolutions, with high memory demands for the explicit
matrix construction~\cite{Grandclement2010-wk, Papenfort:2021hod}, and assuming
symmetry with respect to the orbital plane~\cite{Papenfort:2021hod}.
\texttt{Elliptica} also quote a few days to solve an initial data problem for a
black hole--neutron star binary (BHNS) on 20 cores~\cite{Rashti:2021ihv}.

Despite the time required to solve the elliptic initial data problem,
simulations of merging black holes and neutron stars are currently dominated by
their time evolution, which can take weeks to months. However, significant efforts
are underway to develop faster and more accurate evolution codes for
next-generation numerical relativity. The open-source \spectre{}~\cite{spectre,
Kidder2017-nz} code aims to evolve general-relativistic multiphysics scenarios
on petascale and future exascale computers, and is the main focus of this
article. Other recent developments include the \texttt{CarpetX} driver for the
Einstein Toolkit~\cite{CarpetX} that is based on the \texttt{AMReX}
framework~\cite{AMReX_JOSS}, and the \texttt{Dendro-GR}~\cite{Fernando:2018mov},
\texttt{Nmesh}~\cite{tichy2020numerical}, \texttt{bamps}~\cite{Bugner:2015gqa},
\texttt{GRAthena++}~\cite{Daszuta:2021ecf}, and
\texttt{ExaHyPE}~\cite{Reinarz:2019:ExaHyPE} codes.

To seed these next-generation evolutions of general-relativistic scenarios with
initial data, we have developed a highly scalable elliptic solver based on
discontinuous Galerkin methods, matrix-free iterative algorithms, and task-based
parallelism. We focus strongly on parallelization to take advantage of the
increasing number of cores in high-performance computing (HPC) systems.
These systems have at least $\mathcal{O}(10)$, but often
closer to 50--100, physical cores per node, often with many thousand
interconnected nodes. Therefore, even routine compute jobs that request only a
few nodes on contemporary HPC clusters, and hence spend little to no time waiting in a
queue, have tens to hundreds of cores at their disposal. Larger compute jobs
with thousands of cores and more are also readily available, and the amount of
available computational resources is expected to increase rapidly in the future.

The \spectre{} code embraces parallelism as a core design principle~\cite{spectre,
Kidder2017-nz}. It employs a task-based parallelism paradigm instead of the
conventional message passing interface (MPI) protocol.
MPI-parallelized programs typically alternate between computation and
communication at global synchronization points, meaning that all threads must
reach a globally agreed-upon state before the program proceeds. Global
synchronization points can limit the effective use of the available cores
when some threads reach the synchronization later than others, thus holding
up the program. The effect becomes more pronounced with increasing core count,
often limiting the number of cores that MPI-parallelized programs can
efficiently scale to. Task-based parallel programs, on the other hand, aim to
avoid global synchronization points as much as possible. They partition the
computational work into interdependent tasks and distribute them among the
available cores. Tasks can migrate to undersubscribed cores while the program is
running to balance the computational load. \spectre{} builds upon the
\texttt{Charm++}~\cite{charmpp} task-based parallelism and CPU-abstraction
library. \Ccite{Kidder2017-nz} describes \spectre{}'s task-based parallism paradigm in
more detail.

Our new elliptic solver in the \spectre{} code is based on the prototype
presented in \ccite{Vincent2019qpd} and employs the discontinuous Galerkin
discretization for generic elliptic equations developed
in~\ccite{dgscheme}, which makes it applicable to a wide range of elliptic
problems in numerical relativity and beyond. In this article we present the
task-based iterative algorithms that we have developed to parallelize the
elliptic solver effectively on computing clusters, including novel
subdomain-preconditioning techniques.
We demonstrate that our new elliptic solver can solve a classic initial data
problem for binary black holes faster than \spec{}
when running on as few as eight cores, and in a fraction of the time on a
computing cluster. In particular, the number of iterations that our new elliptic
solver requires to converge remains constant with increasing
resolution. The additional computational work needed to solve high-resolution
problems manifests in subproblems that become either more numerous or more
expensive, but that can be solved in parallel to offset the increase in runtime.

This article is structured as follows. \Cref{sec:discretization} summarizes the
discontinuous Galerkin scheme that was presented in \ccite{dgscheme} and
that we employ to discretize all elliptic equations in this article.
\Cref{sec:algs} details the stack of task-based algorithms that
constitutes the elliptic solver, and that we have implemented in the \spectre{}
code. In \cref{sec:tests} we assess the performance and parallel efficiency of
our new elliptic solver by applying it to a set of test problems. We conclude in
\cref{sec:conclusion}.

\section{Discontinuous Galerkin discretization}\label{sec:discretization}

We employ the discontinuous Galerkin (DG) scheme developed in \ccite{dgscheme}
to discretize all elliptic problems in this article and summarize it in this
section.

Schematically, the discretization procedure translates a linear elliptic problem
to a matrix equation, such as
\begin{equation}\label{eq:matrix_eq}%
  -\partial_i\partial_i \varphi(\bm{x}) = 4\pi\rho(\bm{x})
  \quad \xrightarrow[\text{\ccite{dgscheme}}]{} \quad
  \opA \grd{u} = \grd{\opb}
  \text{,}
\end{equation}
where $\grd{u}=(u_1,\ldots,u_{N_\mathrm{DOF}})$ is a discrete
representation of all variables on the computational grid,
$\grd{\dgb}=(b_1,\ldots,b_{N_\mathrm{DOF}})$~is a discrete representation
of the fixed sources in the PDEs, and~$\dgA$ is an $N_\mathrm{DOF} \times
N_\mathrm{DOF}$ matrix that represents the discrete Laplacian operator in this
example. \Cref{eq:matrix_eq} represents the Maxwell constraint equation for the
electric potential~$\varphi(\bm{x})$ in Coulomb gauge, written here in Cartesian
coordinates, where $\rho(\bm{x})$ is the electric charge density sourcing the
field. We employ the Einstein sum convention to sum over repeated indices.

The subject of this section is to define the matrix
equation~\eqref{eq:matrix_eq} for a wide range of elliptic problems, as detailed
in \ccite{dgscheme}. Then, the remainder of this article is concerned with
solving the matrix equation numerically for~$\grd{u}$, and doing so
iteratively, in parallel on computing clusters, and without ever
explicitly constructing the full matrix~$\dgA$. Instead, we only need to define
the matrix-vector product~$\opA \grd{u}$. We solve nonlinear problems
$\opA(\grd{u})=\grd{b}$ by repeatedly solving their linearization.

The discontinuous Galerkin scheme detailed in \ccite{dgscheme} applies to a wide
range of elliptic problems. Specifically, it applies to any set of elliptic PDEs
that admits a formulation in first-order flux form
\begin{equation}\label{eq:fluxform}
  -\partial_i \, \dgFi{\alpha}[u_A,v_A;\bm{x}] + \dgS_\alpha[u_A,v_A;\bm{x}] = \dgf_\alpha(\bm{x})
  \text{,}
\end{equation}
where the fluxes~$\dgFi{\alpha}$ and the sources~$\dgS_\alpha$ are functionals
of a set of \emph{primal} variables~$u_A(\bm{x})$ and \emph{auxiliary}
variables~$v_A(\bm{x})$, and the fixed sources~$\dgf_\alpha(\bm{x})$ are
functions of coordinates. The index~$\alpha$ enumerates both primal and
auxiliary equations. The primal variables can be scalars, such as the electric
potential~$\varphi(\bm{x})$ in the Maxwell constraint~\eqref{eq:matrix_eq},
higher-rank tensor fields such as the displacement vector in an elasticity
problem, or combinations thereof such as in \cref{eq:xcts} below. The auxiliary
variables are typically gradients of the primal variables, such as
$v_i=\partial_i \varphi(\bm{x})$ for the Maxwell constraint. For example, the
Maxwell constraint~\eqref{eq:matrix_eq} can be formulated with the fluxes and
sources
\begin{subequations}\label{eq:poisson_fluxform}
\begin{alignat}{5}
  \label{eq:poisson_fluxform_aux}
  \tensor{\dgF}{_{\!v}^i_j} &= \varphi \, \delta^i_j
  \text{,} &\quad
  \tensor{\dgS}{_v_{\,j}} &= v_j
  \text{,} &\quad
  \tensor{\dgf}{_v_{\,j}} &= 0
  \text{,} \\
  \label{eq:poisson_fluxform_prim}
  \dgFi{\varphi} &= v_i
  \text{,} &\quad
  \dgS_\varphi &= 0
  \text{,} &\quad
  \dgf_\varphi &= 4\pi\rho(\bm{x})
  \text{,}
\end{alignat}
\end{subequations}
where $\delta^i_j$ denotes the Kronecker delta. Note that
\cref{eq:poisson_fluxform_aux} is the definition of the auxiliary variable, and
\cref{eq:poisson_fluxform_prim} is the Maxwell constraint~\eqref{eq:matrix_eq}.

In particular, the flux form~\eqref{eq:fluxform} also encompasses the extended
conformal thin-sandwich (XCTS) formulation of the Einstein constraint
equations~\cite{Pfeiffer2004-oo},
\begin{subequations}\label{eq:xcts}
\begin{align}
  \label{eq:xcts_hamiltonian}
  \bar{\nabla}^2 \psi &= \begin{aligned}[t]
    &\frac{1}{8}\psi\bar{R} + \frac{1}{12}\psi^5 K^2 \\
    &-\frac{1}{8}\psi^{-7}\bar{A}_{ij}\bar{A}^{ij} - 2\pi\psi^5\rho
  \end{aligned}
  \\
  \label{eq:xcts_lapse}
  \bar{\nabla}^2\left(\alpha\psi\right) &= \alpha\psi
  \begin{aligned}[t]
    &\bigg(\frac{7}{8}\psi^{-8}\bar{A}_{ij}\bar{A}^{ij} + \frac{5}{12}\psi^4 K^2 + \frac{1}{8}\bar{R} \\
    &+ 2\pi\psi^4\left(\rho + 2S\right)\bigg) - \psi^5\partial_t K + \psi^5\beta^i\bar{\nabla}_i K
  \end{aligned}
  \\
  \label{eq:xcts_momentum}
  \bar{\nabla}_i(\bar{L}\beta)^{ij} &= \begin{aligned}[t]
    &(\bar{L}\beta)^{ij}\bar{\nabla}_i \ln(\bar{\alpha}) + \bar{\alpha}\bar{\nabla}_i\left(\bar{\alpha}^{-1}\bar{u}^{ij}\right) \\
    &+ \frac{4}{3}\bar{\alpha}\psi^6\bar{\nabla}^j K + 16\pi\bar{\alpha}\psi^{10}S^j
  \end{aligned}
\end{align}
\end{subequations}
with $\bar{\nabla}^2 = \bar{\gamma}^{ij} \bar{\nabla}_i \bar{\nabla}_j$,
$\bar{A}^{ij} = \frac{1}{2\bar{\alpha}}\left((\bar{L}\beta)^{ij} -
\bar{u}^{ij}\right)$ and $\bar{\alpha} = \alpha \psi^{-6}$. The XCTS equations
are a set of coupled nonlinear elliptic PDEs that the spacetime metric of
general relativity must satisfy at all times.\footnote{See, e.g.,
\ccite{BaumgarteShapiro} for an introduction to the XCTS equations, in
particular Box~3.3.} They are solved for the conformal factor~$\psi$, the
product of lapse and conformal factor~$\alpha\psi$, and the shift
vector~$\beta^j$. The remaining quantities in the equations, i.e., the conformal
metric~$\bar{\gamma}_{ij}$, the trace of the extrinsic curvature~$K$, their
respective time derivatives $\bar{u}_{ij}$ and $\partial_t K$, the energy
density~$\rho$, the stress-energy trace~$S$ and the momentum density~$S^i$, are
freely-specifiable fields that define the scenario at hand. In particular, the
conformal metric~$\bar{\gamma}_{ij}$ defines the background geometry of the
elliptic problem, which determines the covariant derivative~$\bar{\nabla}$, the
Ricci scalar~$\bar{R}$ and the longitudinal operator
\begin{equation}
  \left(\bar{L}\beta\right)^{ij} = \bar{\nabla}^i\beta^j + \bar{\nabla}^j\beta^i
  - \frac{2}{3}\bar{\gamma}^{ij}\bar{\nabla}_k\beta^k
  \text{.}
\end{equation}
\Ccite{dgscheme} lists fluxes and sources for the XCTS equations, and for a
selection of other elliptic systems.

\begin{figure}
  \centering
  \subfloat[Domain\protect\label{fig:domain}]{
    \tikzinput{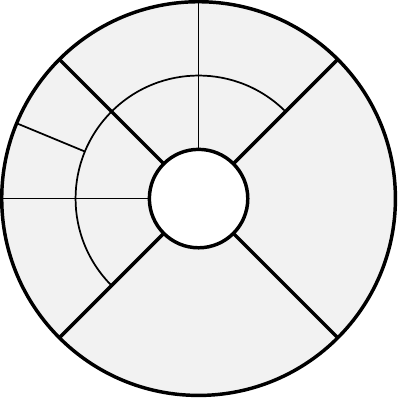}
  }\\
  \subfloat[Block\protect\label{fig:block}]{
    \tikzinput{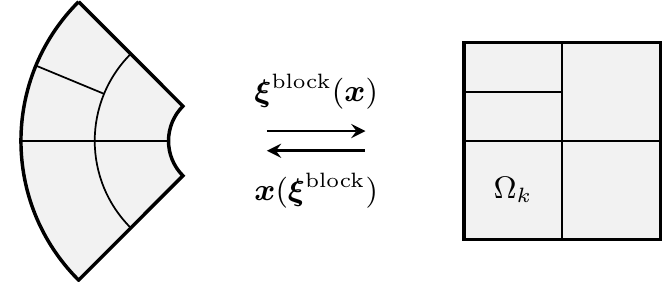}
  }\\
  \subfloat[Element $\Omega_k$\protect\label{fig:element}]{
    \tikzinput{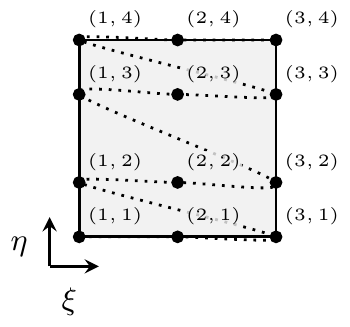}
  }
  \caption{
    \label{fig:domain_and_elements}
    \emph{Top:} Geometry of a two-dimensional computational domain composed of
    four wedge-shaped blocks. \emph{Middle:} The coordinate
    transformation~$\bm{\xi}^\mathrm{block}(\bm{x})$ maps a block to a reference
    cube in block-logical coordinates~$[-1,1]^2$. A block is split into
    elements~$\Omega_k$ along its logical coordinates axes. \emph{Bottom:} The
    element $\Omega_k$ in element-logical coordinates~$\bm{\xi}=(\xi,\eta)$ with
    its grid of Legendre-Gauss-Lobatto collocation points. In this example we
    chose $N_{k,\xi}=3$ and $N_{k,\eta}=4$. Each grid point is labeled with its
    index~$(p_\xi,p_\eta)$. The dotted line connects points in the order they
    are enumerated in by the index~$p$.}
\end{figure}

Once we have formulated the equations, we choose a computational domain on which
to discretize them. We decompose the $d$-dimensional computational domain
$\Omega \subset \mathbb{R}^d$ into a set of \emph{blocks} shaped like deformed
cubes, as illustrated in \cref{fig:domain}. Blocks do not overlap, but they
share boundaries. Each face of a block is either shared with precisely one other
block, or is external. For example, the domain depicted in \cref{fig:domain} has
four wedge-shaped blocks. Each block $B \subset \Omega$ carries a map from the
coordinates $\bm{x} \in B$, in which the elliptic equations~\eqref{eq:fluxform}
are formulated, to \emph{block-logical} coordinates $\bm{\xi}^\mathrm{block} \in
[-1, 1]^d$ representing the $d$-dimensional reference cube, as illustrated in
\cref{fig:block}.

Blocks decompose into \emph{elements}~$\Omega_k \subset \Omega$, by recursively
splitting in half along any of their logical coordinate axes ($h$~refinement). We
limit the $h$~refinement of our computational domain such that an element shares
its boundary with at most two neighbors per dimension in every direction
(\enquote{two-to-one balance}), both within a block and across block boundaries.
Each element defines element-logical coordinates $\bm{\xi} \in [-1, 1]^d$ by an
affine transformation of the block-logical coordinates. The resulting coordinate
map to the reference cube of the element is characterized by the Jacobian
\begin{equation}
  \jac^i_j = \pdv{{x^i}}{{\xi^j}}
\end{equation}
with determinant~$\jac$ and inverse~$\invjac^j_i = \partial \xi^j / \partial
x^i$.

On the reference cube of the element we choose a regular grid of collocation
points along the logical coordinate axes, as illustrated in \cref{fig:element}
($p$~refinement). Specifically, we choose $N_{k, i}$ Legendre-Gauss-Lobatto (LGL)
collocation points, $\xi_{p_i}$, in each dimension~$i$, where the index
$p_i\in\{1,\ldots,N_{k,i}\}$ identifies the grid point along dimension~$i$. We
also enumerate all $N_k=\prod_i N_{k, i}$ $d$-dimensional grid
points~$\bm{\xi}_p=(\xi_{p_1},\ldots,\xi_{p_d})$ in an element with a single
index $p\in\{1,\ldots,N_k\}$, as illustrated in \cref{fig:element}.

Then, fields are represented numerically by their values at the collocation
points. We denote the set of discrete field values within an element~$\Omega_k$
as $\grd{u}^{(k)}=(u_1,\ldots,u_{N_k})$, and the collection of discrete
field values over \emph{all} elements as~$\grd{u}$. The field values at
the collocation points within an element define a $d$-dimensional Lagrange
interpolation,
\begin{equation}\label{eq:field_expansion}
  u^{(k)}(\bm{x}) \defeq \sum_{p=1}^{N_k} u_p \bas_p(\bm{\xi}(\bm{x}))
  \quad \text{with} \quad \bm{x} \in \Omega_k
  \text{,}
\end{equation}
where the basis functions~$\bas_p(\bm{\xi})$ are products of Lagrange
polynomials,
\begin{equation}\label{eq:lagrprod}
  \bas_p(\bm{\xi}) \defeq \prod_{i=1}^d \lagr_{p_i}(\xi^i)
  \quad \text{with} \quad \bm{\xi} \in [-1,1]^d
  \text{.}
\end{equation}
based on the collocation points in dimension~$i$ of the element. Since
\cref{eq:field_expansion,eq:lagrprod} are local to each element, fields
over the entire domain are discontinuous across element boundaries.

Finally, we employ the strong discontinuous Galerkin scheme
developed in \ccite{dgscheme} to discretize the equations in first-order flux
form, \cref{eq:fluxform}. To compute the matrix-vector product in
\cref{eq:matrix_eq} we first compute the auxiliary variables $v_A$, given the
primal variables $u_A$, as
\begin{subequations}\label{eq:dg_residuals}
\begin{equation}\label{eq:dg_op_auxvars}
  v_A = \dgD_i \cdot \dgFi{v_A}
  + \dgL \cdot ((n_i\dgFi{v_A})^* - n_i\dgFi{v_A})
  - \tilde{\dgS}_{v_A}
  \text{,}
\end{equation}
where we assume the auxiliary sources can be written in the form $\dgS_{v_A}=v_A
+ \tilde{\dgS}_{v_A}[u_A;\bm{x}]$ such that \cref{eq:dg_op_auxvars} depends
only on the primal variables. We also assume $\dgf_{v_A}=0$ for convenience. All
elliptic equations that we consider in this article fulfill these assumptions.
In a second step, we use the computed auxiliary variables $v_A$, as well as the
primal variables $u_A$, to compute the DG residuals
\begin{align}\label{eq:dg_op_primal}
  -\dgMD_i \cdot \dgFi{u_A}
  - \dgML \cdot ((n_i \dgFi{u_A})^* &- n_i \dgFi{u_A}) \nonumber \\
  &+ \dgM \cdot \dgS_{u_A}
  = \dgM \cdot \dgf_{u_A}
  \text{.}
\end{align}
\end{subequations}
The operation~$\cdot$ in \cref{eq:dg_residuals} denotes a matrix multiplication
with the field values over the computational grid of an element. We make use of
the mass matrix
\begin{align}
  \label{eq:massmat}
  \dgM_{pq} &= \int_{[-1,1]^d} \bas_p(\bm{\xi}) \bas_q(\bm{\xi}) \sqrt{g} \, \jac \dd{^d\xi}
  \text{,}
\intertext{the stiffness matrix}
  \label{eq:stiffmat}
  \dgMD_{i,pq} &= \int_{[-1,1]^d} \bas_p(\bm{\xi}) \, \pdv{\bas_q}{{\xi^j}}\!(\bm{\xi})
  \, \invjac^j_i \sqrt{g} \, \jac \dd{^d\xi}
  \text{,}
\intertext{and the lifting operator}
  \label{eq:liftop}
  \dgML_{pq} &= \int_{[-1,1]^{d-1}} \bas_p(\bm{\xi})\bas_q(\bm{\xi})
  \sqrt{\surf{g}} \, \surf{\jac} \dd{^{d-1}\xi}
\end{align}
on the element~$\Omega_k$, as well as the associated "massless" operators
$\dgD_i \defeq \dgM^{-1}\dgMD_i$ and $\dgL \defeq \dgM^{-1}\dgML$. Here,
$\sqrt{g}$ denotes the determinant of the metric on which the elliptic equations
are formulated, such as the conformal metric~$\bar{\gamma}_{ij}$ in the XCTS
equations~\eqref{eq:xcts}. The integral in \cref{eq:liftop} is over the boundary
of the element, $\partial\Omega_k$, where $n_i$ is the outward-pointing unit
normal one-form, $\surf{g}$ is the surface metric determinant induced by the
background metric, and $\surf{\jac}$ is the determinant of the surface Jacobian.

The quantities $(n_i \dgFi{v_A})^*$ and $(n_i \dgFi{u_A})^*$ in
\cref{eq:dg_residuals} denote a numerical flux that couples grid points across
nearest-neighbor element boundaries. We employ the generalized internal-penalty
numerical flux developed in \ccite{dgscheme},
\begin{subequations}\label{eq:numflux}
\begin{align}
  (n_i \dgFi{v_A})^* = \frac{1}{2}\Bigl[
    &n_i^\mathrm{int} \dgFi{v_A}(u_A^\mathrm{int}) -
    n_i^\mathrm{ext} \dgFi{v_A}(u_A^\mathrm{ext})\Bigr]
    \label{eq:numfluxaux}\text{,}\\
  (n_i \dgFi{u_A})^* =
    \frac{1}{2}\Bigl[
    &n_i^\mathrm{int} \dgFi{u_A}\bigl(\partial_j \dgFj{v_A}(u_A^\mathrm{int}) - \tilde{\dgS}_{v_A}(u_A^\mathrm{int})\bigr) \nonumber \\
    &- n_i^\mathrm{ext} \dgFi{u_A}\bigl(\partial_j \dgFj{v_A}(u_A^\mathrm{ext}) - \tilde{\dgS}_{v_A}(u_A^\mathrm{ext})\bigr)\Bigr] \nonumber \\
    - \sigma \Bigl[
      &n_i^\mathrm{int} \dgFi{u_A}\bigl(n_j^\mathrm{int} \dgFj{v_A}(u_A^\mathrm{int})\bigr) \nonumber \\
      &- n_i^\mathrm{ext} \dgFi{u_A}\bigl(n_j^\mathrm{ext} \dgFj{v_A}(u_A^\mathrm{ext})\bigr)\Bigr] \label{eq:numfluxprimal}
\end{align}
\end{subequations}
with the penalty function
\begin{equation}\label{eq:penalty}
  \sigma = \penaltyparam \, \frac{\big(\max(p^\mathrm{int},p^\mathrm{ext})+1\big)^2}{\min(h^\mathrm{int}, h^\mathrm{ext})}
  \text{.}
\end{equation}
Here, $u_A^\mathrm{int}$ denotes the primal variables on the \emph{interior}
side of an element's shared boundary with a neighbor, and $u_A^\mathrm{ext}$
denotes the primal variables on the neighbor's side, i.e., the \emph{exterior}.
Note that $n_i^\mathrm{ext} = -n_i^\mathrm{int}$ for the purpose of this
article, since we only consider equations formulated on a fixed background
metric, but the scheme does not rely on this assumption. For \cref{eq:penalty}
we also make use of the polynomial degree~$p$, and a measure of the element
size, $h$, orthogonal to the element boundary on either side of the interface,
as detailed in \ccite{dgscheme}.

We impose boundary conditions through fluxes, i.e., by a choice of exterior
quantities in the numerical flux, \cref{eq:numflux}. Specifically, on external
boundaries we set
\begin{equation}\label{eq:bc}
  (n_i\dgFi{\alpha})^\mathrm{ext} = (n_i\dgFi{\alpha})^\mathrm{int} - 2 (n_i\dgFi{\alpha})^\mathrm{b}
  \text{,}
\end{equation}
where we choose the boundary fluxes $(n_i\dgFi{\alpha})^\mathrm{b}$ depending on
the boundary conditions we intend to impose. For \emph{Neumann-type} boundary
conditions we choose the primal boundary fluxes $(n_i\dgFi{u_A})^\mathrm{b}$
directly, e.g., $(n_i\dgFi{\varphi})^\mathrm{b} =
n_i\partial_i\varphi|_\mathrm{b}$ for the Maxwell
constraint~\eqref{eq:matrix_eq}, and set the auxiliary boundary fluxes to their
interior values, $(n_i\dgFi{v_A})^\mathrm{b}=(n_i\dgFi{v_A})^\mathrm{int}$. For
\emph{Dirichlet-type} boundary conditions we choose the primal boundary
fields~$u_A^\mathrm{b}$, e.g., $\varphi|_\mathrm{b}$ for the Maxwell
constraint~\eqref{eq:matrix_eq}, to compute the auxiliary boundary fluxes
$(n_i\dgFi{v_A})^\mathrm{b} = n_i^\mathrm{b}\dgFi{v_A}(u_A^\mathrm{b})$, and set
the primal boundary fluxes to their interior values,
$(n_i\dgFi{u_A})^\mathrm{b}=(n_i\dgFi{u_A})^\mathrm{int}$.

In summary, the DG residuals~\eqref{eq:dg_residuals} are algebraic equations for
the discrete primal field values $\grd{u}_A$ on all elements and grid
points in the computational domain. For linear PDEs, the left-hand side of
\cref{eq:dg_op_primal} defines a matrix-vector product with a set of primal
field values on the computational domain. The right-hand side of
\cref{eq:dg_op_primal} is a set of fixed values on the computational domain.
Therefore, \cref{eq:dg_op_primal} has the form of \cref{eq:matrix_eq}.

\section{Task-based iterative algorithms}\label{sec:algs}

Once the elliptic problem is discretized, it is the responsibility of the
elliptic solver to invert the matrix equation~\eqref{eq:matrix_eq} numerically,
in order to obtain the solution vector~$\grd{u}$ over the computational grid.
For large problems on high-resolution grids it is typically unfeasible to invert
the matrix~$\opA$ in \cref{eq:matrix_eq} directly, or even to explicitly
construct and store it.\footnote{The \texttt{KADATH}~\cite{Grandclement2010-wk}
code explicitly constructs, stores and inverts the matrix~$\opA$, which it
obtains by a spectral discretization, by distributing its columns over the
available cores. \Ccites{Grandclement2010-wk,Papenfort:2021hod} quote the high
memory demand of storing the explicit matrix and list iterative approaches to
solve the linear system as a possible resolution. Such iterative approaches, and
their parallelization, are the main focus of this article.} Instead, we employ
iterative algorithms that require only the matrix-vector product $\opA\grd{u}$
be defined, and that parallelize to computing clusters.

\begin{figure}
  \centering
  \subfloat[Domain]{
    \tikzinput{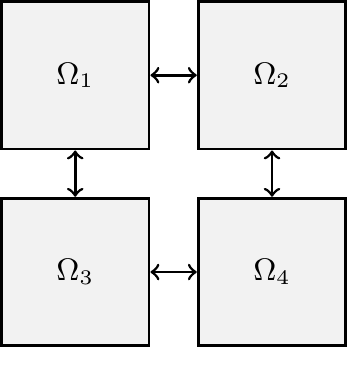}
  }\\
  \subfloat[Tasks]{
    \tikzinput{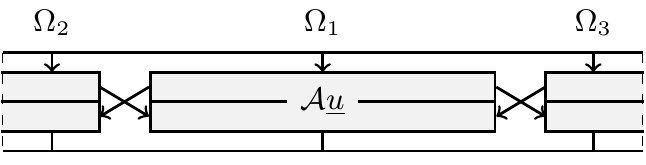}
  }
  \caption{
    \label{fig:parallel_operator}
    Parallelization structure of the matrix-vector product~$\opA \grd{u}$.
    \emph{Top:} Decomposition of a two-dimensional rectangular domain into four
    elements. Arrows illustrate the dependence between nearest-neighbor elements
    to compute the matrix-vector product~$\opA \grd{u}$. \emph{Bottom:} Tasks
    involved to compute the matrix-vector product~$\opA \grd{u}$. Each element
    performs a task that prepares and sends data to its neighbors (upper half of
    the rectangle), and another that receives data from its neighbors and
    performs the computation (lower half of the rectangle). The arrows between
    elements are the same as in the top panel.}
\end{figure}

The discontinuous Galerkin (DG) matrix-vector product~$\opA\grd{u}$ is
well suited for parallelization. As \cref{sec:discretization} summarized,
it decomposes into a set of
operations local to the elements that make up the computational domain.
\Cref{fig:parallel_operator} illustrates a computational domain composed of
elements, as well as the dependence of the elements on each other for computing
the matrix-vector product. The matrix-vector product requires only data local to
each element and on both sides of the boundary that the element shares with its
nearest neighbors. Therefore, it can be computed in parallel, and requires only
a single communication between each pair of nearest-neighbor elements to
exchange data on their shared boundary. The matrix-vector product acts as a
\enquote{soft} global synchronization point, meaning that it requires \emph{all}
elements have sent data to their neighbors before \emph{all} elements can
proceed with the algorithm, but individual elements can already proceed once
they receive data from their nearest neighbors.

The decomposition of the domain into elements also admits a strategy to
distribute computation across the processors of the computer system. We
distribute elements among the available cores in a way that, ideally, minimizes
the number of internode communications and assigns an equal amount of work to
each core. In this article we employ a Morton (\enquote{z-order}) space-filling
curve~\cite{Sagan1994} to traverse the elements within a block of the
computational domain and fill up the available cores. We weight the elements by
their number of grid
points to approximately balance the amount of work assigned to each core. With
this strategy, neighboring elements tend to lie on the same node, though more
effective element-distribution and load-balancing strategies based on, for
example, Hilbert space-filling curves~\cite{Borrell2018} are a subject of future work.

Once elements have been assigned to the available cores, each element traverses
the list of tasks in the algorithm. When it encounters a task whose dependencies
are not yet fulfilled, e.g., when neighbors have not yet sent the data on shared
boundaries needed for the DG matrix-vector product, the element relinquishes
control of the core to another whose dependencies are fulfilled.
\Ccite{Kidder2017-nz} describes \spectre{}'s task-based parallel runtime system
based on the \texttt{Charm++}~\cite{charmpp} framework in more detail.

\begin{figure}
  \centering
  \tikzinput{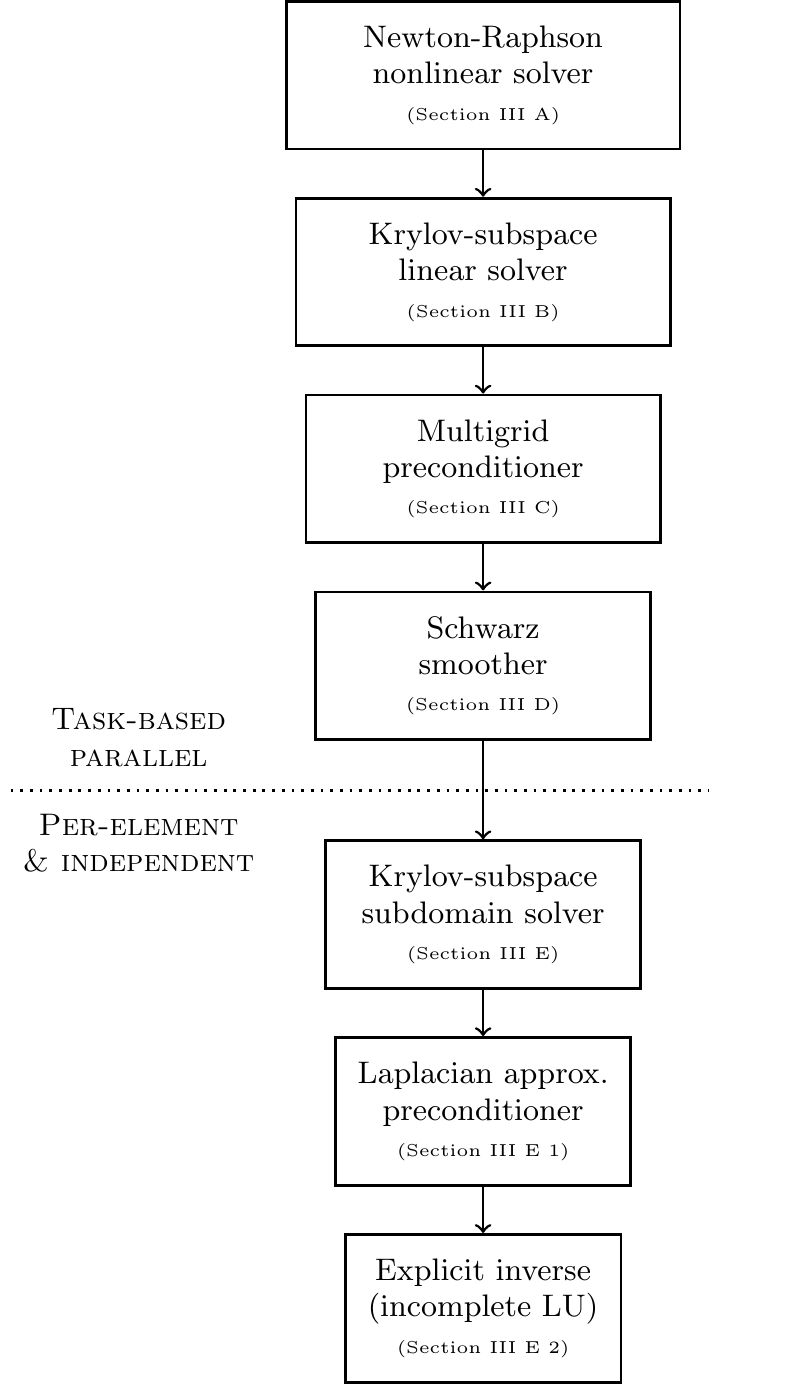}
  \caption{
    \label{fig:overview}
    Overview of the technology stack we employ to solve the discretized elliptic
    problem~\eqref{eq:matrix_eq}. All algorithms above the dotted line follow
    \spectre{}'s task-based parallelism paradigm. The algorithms below the
    dotted line run within a task, and on all elements independently.}
\end{figure}

\Cref{fig:overview} provides an overview of the algorithms that we employ to
iteratively solve the discretized elliptic problem~\eqref{eq:matrix_eq}, with
details given in subsequent sections: nonlinear equations are linearized with a
Newton-Raphson scheme with a line-search globalization (\cref{sec:nonlin}). The
resulting linear subproblems are solved with an iterative Krylov-subspace method
(\cref{sec:lin}), preconditioned with a multigrid solver (\cref{sec:mg}). On
every level of the multigrid hierarchy we run a few iterations of an additive
Schwarz smoother, which solves the problem approximately on independent,
overlapping, element-centered subdomains (\cref{sec:schwarz}). Each subdomain
problem is solved by another Krylov-type method, which carries a
Laplacian-approximation preconditioner with an incomplete LU explicit-inversion
scheme to accelerate the solve (\cref{sec:subd}). All algorithms are implemented
in the open-source \spectre{} code and take advantage of its task-based parallel
infrastructure~\cite{spectre, Kidder2017-nz}.

\subsection{Newton-Raphson nonlinear solver}\label{sec:nonlin}

The Newton-Raphson scheme iteratively refines an initial guess $\grd{u}_0$ for a
nonlinear problem $\opA(\grd{u}) = \grd{b}$ by repeatedly solving the
linearized problem
\begin{equation}\label{eq:linearized}
\frac{\delta \opA}{\delta u}(\grd{u}) \, \Delta \grd{u} =
\grd{b}-\opA(\grd{u})
\end{equation}
for the correction $\Delta \grd{u}$, and then updating the solution as
$\grd{u} \rightarrow \grd{u} + \Delta \grd{u}$~\cite{numericalrecipes,
DennisSchnabel}.

The Newton-Raphson method converges quadratically once it reaches a basin of
attraction, but can fail to converge when the initial guess is too far from the
solution. We employ a line-search globalization strategy to recover convergence
in such cases, following Alg.~6.1.3 in \ccite{DennisSchnabel}. It iteratively
reduces the step length~$\lambda$ until the corrected residual $\lVert\grd{b}
- \opA(\grd{u} + \lambda \Delta \grd{u})\rVert_2$ has sufficiently
decreased, meaning it has decreased by a fraction of the predicted decrease if
the problem was linear. This fraction is the \emph{sufficient-decrease
parameter} controlling the line search. The line search typically starts at
${\lambda=1}$ in every Newton-Raphson iteration, but the initial step length can
be decreased to dampen the
nonlinear solver. Although the line-search globalization has proven effective
for the cases we have encountered so far, alternative globalization strategies
such as a trust-region method or more sophisticated nonlinear preconditioning
techniques can be investigated in the future.\footnote{See, e.g.,
\ccite{Brune2015} for an overview of nonlinear preconditioning techniques in
the context of the \texttt{PETSc}~\cite{petsc} library.}

\begin{figure}
  \centering
  \tikzinput{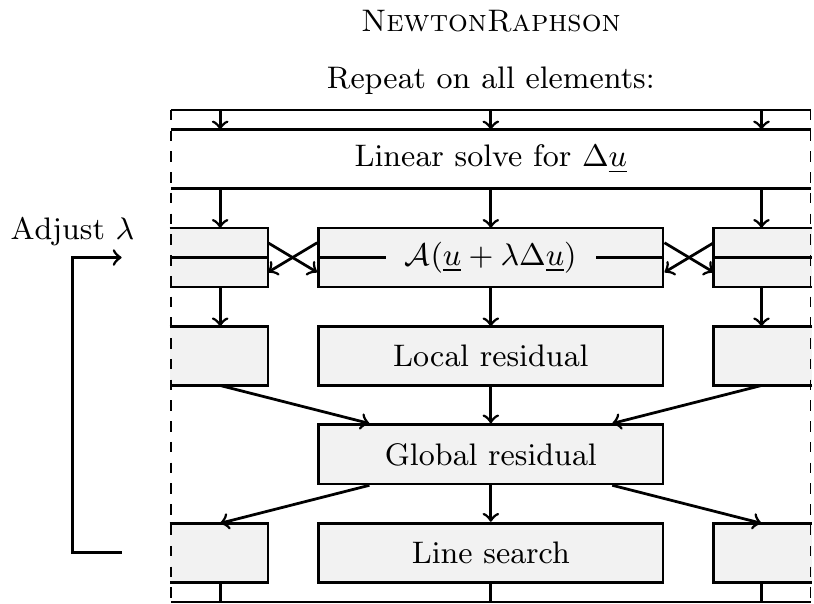}
  \caption{
    \label{fig:nonlin_tasks}
    Parallelization structure of the task-based Newton-Raphson nonlinear
    solver (\cref{sec:nonlin}).}
\end{figure}

\Cref{fig:nonlin_tasks} illustrates our task-based implementation of the
Newton-Raphson algorithm. The sufficient-decrease condition, and the necessity
to check the global residual magnitude against convergence criteria, introduce a
synchronization point in the form of a global reduction to assemble the residual
magnitude~$\lVert\grd{b} - \opA(\grd{u} + \lambda\Delta\grd{u})\rVert_2$.
The algorithm requires one nonlinear operator application~$\opA(\grd{u} +
\lambda \Delta\grd{u})$ per iteration, plus one additional nonlinear operator
application for every globalization step that reduces the step length. Since a
typical nonlinear elliptic solve requires $\lesssim 10$
Newton-Raphson iterations, the parallelization properties of this algorithm are
not particularly important for the overall performance.

Exactly once per iteration the Newton-Raphson algorithm dispatches a linear
solve of \cref{eq:linearized} for the correction~$\Delta \grd{u}$. This
iterative linear-solver algorithm is the subject of the following section.

\subsection{Krylov-subspace linear solver}\label{sec:lin}

We solve the linearized problem~\eqref{eq:linearized} with an iterative
Krylov-subspace algorithm. We generally employ a GMRES algorithm, but have also
developed a conjugate gradients algorithm for discretized problems that are
symmetric positive definite~\cite{Saad1986-ix,Saad2003}. These algorithms solve
a linear problem $\opA\grd{u}=\grd{b}$ iteratively by building up a basis of the
Krylov subspace $K_k=\left\{\grd{b}, \opA\grd{b}, \opA^2\grd{b}, \ldots,
\opA^{k-1}\grd{b}\right\}$. Krylov-subspace algorithms are guaranteed to find a
solution in at most $N_\mathrm{DOF}$ iterations, where $N_\mathrm{DOF}$ is the
size of the matrix~$\opA$.

\begin{figure}
  \centering
  \tikzinput{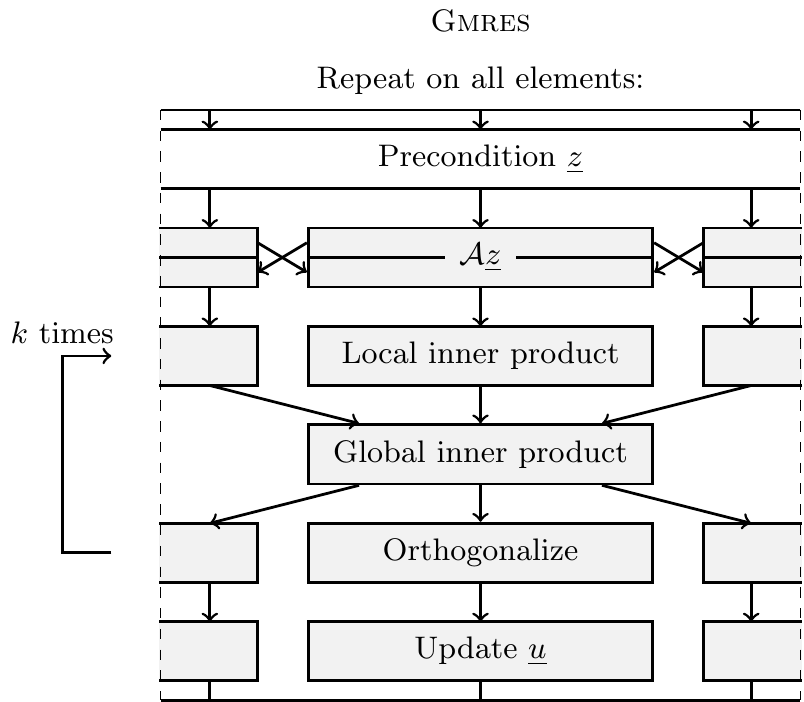}
  \caption{
    \label{fig:gmres_tasks}
    Parallelization structure of the task-based GMRES Krylov-subspace linear
    solver (\cref{sec:lin}).}
\end{figure}

\Cref{fig:gmres_tasks} illustrates our task-based GMRES algorithm, which is
based on Alg.~9.6 in \ccite{Saad2003}. It requires one application of the
linear operator~$\opA$ per iteration. Then, the algorithm is characterized by an
Arnoldi orthogonalization procedure to construct a new basis vector $\grd{z}$ of the
Krylov subspace that is orthogonal to all previously constructed basis vectors.
The orthogonalization procedure requires a global reduction to assemble the
inner product of the new basis vector with every existing basis vector, meaning
the GMRES algorithm needs to perform $k$ reductions in the $k$th iteration.
Every reduction constitutes a global synchronization point, since it requires
that all elements send data to a single core on the computer system and wait for
a broadcast from that core back to all elements. A conjugate gradients algorithm
also requires a global reduction per iteration, but avoids the additional
reductions from the orthogonalization procedure.

Due to the global synchronization points involved in every iteration of the
Krylov-subspace solver, it is essential to keep the number of iterations to a
minimum in order to achieve good parallel performance. To this end, we invoke a
\emph{preconditioner} in every iteration of the Krylov-subspace algorithm and
place particular focus on its parallelization properties. The preconditioner is
responsible for solving the linear problem approximately to accelerate the
convergence of the Krylov-subspace algorithm.\footnote{See, e.g.,
\ccite{Saad2003} for an introduction to iterative linear solvers and
preconditioning techniques.} Effective parallel preconditioning techniques for
our DG-discretized elliptic problems are the main focus of this article. Since
we employ the \emph{flexible} variant of the GMRES algorithm, the preconditioner
may change between iterations~\cite{Saad2003}.  While the flexible GMRES
algorithm with a variable preconditioner is not mathematically guaranteed to
converge in at most $N_\mathrm{DOF}$ iterations anymore, in practice, it
converges in much fewer than $N_\mathrm{DOF}$ iterations (see test problems in
\cref{sec:tests}, in particular
\cref{fig:poisson_iterations,fig:kerrschild_iterations}).\footnote{See also
Sec.~9.4.1 in \ccite{Saad2003} for a discussion of the flexible GMRES
algorithm.}

Typically, the number of iterations needed by an unpreconditioned
Krylov-subspace algorithm increases with the size of the problem. The
convergence behavior is often connected to the condition number of the linear
operator,
\begin{equation}
  \kappa = \frac{\lambda_\mathrm{max}}{\lambda_\mathrm{min}}
  \text{,}
\end{equation}
where $\lambda_\mathrm{max}$ and $\lambda_\mathrm{min}$ denote the largest and
smallest eigenvalue of the matrix, respectively. However, note that rigorous
convergence bounds for the GMRES algorithm in terms of the condition number
exist only when the matrix is \emph{normal} \cite{Saad2003}. Nevertheless,
the condition number can provide an indication for the expected rate of
convergence. For the discontinuous Galerkin discretization we employ in this
article, the condition number scales as~$\kappa \propto p^2 / h$, where~$p$
denotes a typical polynomial degree of the elements and~$h$ denotes a typical
element size~\cite{dgscheme, Vincent2019qpd, HesthavenWarburton,
Shahbazi2005-dr}. This scaling is related to the decrease of the minimum
spacing between Legendre-Gauss-Lobatto collocation points, which scales
quadratically with $p$ near element boundaries and linearly with the element
size.

More specifically, Krylov-subspace methods struggle to solve large-scale modes
in the solution.
The algorithm solves modes on the scale of the grid-point spacing or the size of
elements in just a few iterations, but it needs a lot more iterations to solve
modes spanning the full domain. Such large-scale modes carry, for example,
information from boundary conditions that must traverse the entire domain.
The test problem presented in \cref{fig:poisson_smoothing} below illustrates
this effect.
Therefore, we precondition the Krylov-subspace solver with a multigrid algorithm
that uses information from coarser grids, where the large-scale modes from finer
grids become small scale.

\subsection{Multigrid preconditioner}\label{sec:mg}

We employ a geometric \vcycle{} multigrid algorithm, as prototyped
in~\ccite{Vincent2019qpd}.\footnote{See, e.g., \ccite{Briggs2000jp} for an
introduction to multigrid methods.} Our multigrid solver can be used
standalone, or to precondition a Krylov-type linear solver
as described in \cref{sec:lin} (\enquote{Krylov-accelerated multigrid}).

\subsubsection{Grid hierarchy}\label{sec:mg_hierarchy}

\begin{figure}
  \centering
  \tikzinput{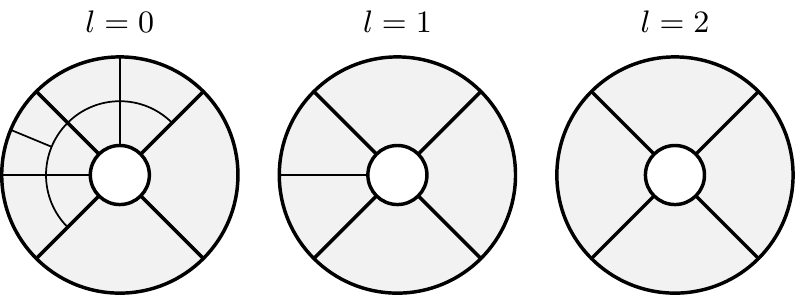}
  \caption{
    \label{fig:mg_hierarchy}
    Multigrid hierarchy based on the domain depicted in \cref{fig:domain}.
    }
\end{figure}

The geometric multigrid algorithm relies on a strategy to coarsen the
computational grid. We primarily $h$-coarsen the domain, meaning that we create multigrid
levels $l > 0$ by successively combining two elements into one along every
dimension of the grid, as illustrated in \cref{fig:mg_hierarchy}. We only
$p$-coarsen the grid in the sense that we choose the smaller of the two polynomial
degrees when combining elements along an axis. This strategy follows
\ccite{Vincent2019qpd} and ensures that coarse-grid field approximations always
have an exact polynomial representation on finer grids.

The coarsest possible grid that our domain decomposition can achieve has a
single element per block that make up the domain. For example, the
two-dimensional shell depicted in \cref{fig:mg_hierarchy} has four wedge-shaped
blocks, each of which is a deformed cube. Our multigrid algorithm works best
when the domain is composed of as few blocks as possible.

\subsubsection{Intermesh operators}

To project data between grids we use the standard $L_2$-projections (or
\emph{Galerkin} projections) detailed in \ccite{dgscheme}.\footnote{See also
\ccite{Fortunato2019} for details on the intermesh operators.} Fields on
coarser grids are projected to finer grids with the
\emph{prolongation operator}
\begin{equation}\label{eq:prol}
  \prol_{\tilde{p} p}^{\, l+1 \rightarrow l} = \prod_{i=1}^d\lagr_{p_i}(\tilde{\xi}_{\tilde{p}_i})
  \text{,}
\end{equation}
where $p$ enumerates grid points on the coarser grid, $\tilde{p}$ enumerates
grid points on the finer grid, and $\tilde{\xi}_{\tilde{p}_i}$ are the
coarse-grid logical coordinates of the fine-grid collocation points. For
fine-grid (child) elements that cover the full coarse-grid (parent) element in
dimension $i$ the coarse-grid logical coordinates are just the fine-grid
collocation points, $\tilde{\xi}_{\tilde{p}_i}=\xi_{\tilde{p}_i}$. For child
elements that cover the lower or upper logical half of the parent element in
dimension~$i$ they are $\tilde{\xi}_{\tilde{p}_i}=(\xi_{\tilde{p}_i}-1)/2$ or
$\tilde{\xi}_{\tilde{p}_i}=(\xi_{\tilde{p}_i}+1)/2$, respectively. Note that the
prolongation operator~\eqref{eq:prol} is just a Lagrange interpolation from the
coarser to the finer grid. The interpolation retains the accuracy of the
polynomial approximation because the finer grid always has sufficient resolution.

To project data from finer to coarser grids we employ the \emph{restriction
operator}
\begin{equation}
  \restr^{\, l \rightarrow l+1} =(\prol^{\, l+1 \rightarrow l})^T
  \text{,}
\end{equation}
which is the transpose of the prolongation operator~\eqref{eq:prol}. Contrary to
the restriction operator listed in~\ccite{dgscheme} the multigrid restriction
involves no mass matrices because it applies to DG residuals,
\cref{eq:dg_op_primal}, which already include mass matrices.

\subsubsection{Algorithm}

\begin{figure*}
  \centering
  \tikzinput{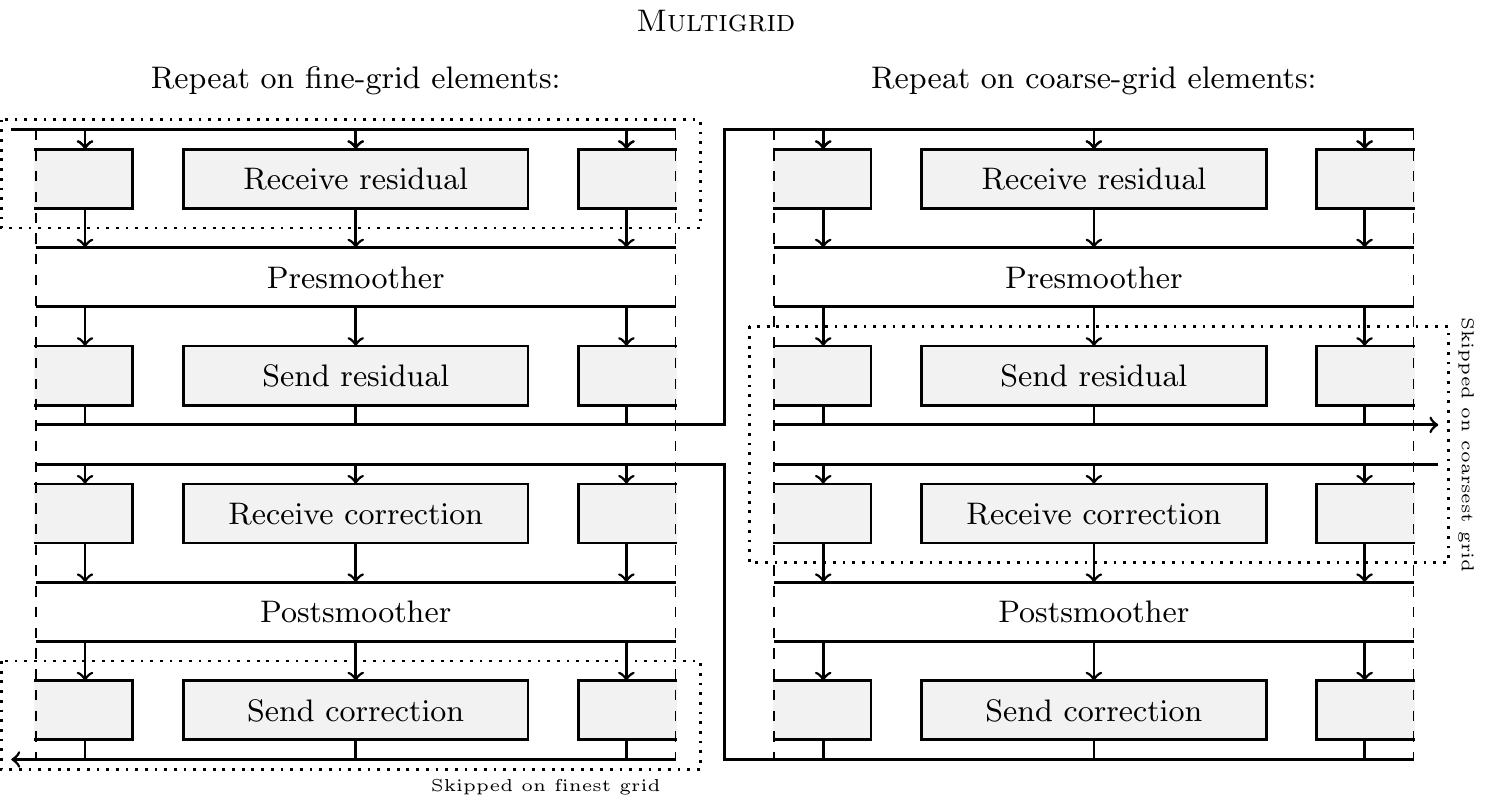}
  \caption{
    \label{fig:mg_tasks}
    Parallelization structure of the task-based multigrid algorithm
    (\cref{sec:mg}). Elements on all grids perform the same set of tasks, with
    some tasks skipped on the finest grid and other tasks skipped on the
    coarsest grid.}
\end{figure*}

\Cref{fig:mg_tasks} illustrates our task-based implementation of the multigrid
\vcycle{} algorithm to solve linear problems $\opA\grd{u} = \grd{b}$.
On every grid~$l$ we approximately solve the linear problem
\begin{equation}\label{eq:mg_problem}
  \opA_l \, \grd{u}^{(l)} = \grd{b}^{(l)}
  \text{,}
\end{equation}
where the operator $\opA_l$ is the discretization of the elliptic PDEs on the
grid $l$. At the beginning of a \vcycle{}, on the finest grid $l=0$, we select
$\grd{u}^{(0)} = \grd{u}$ and $\grd{b}^{(0)} = \grd{b}$; hence, approximately
solving the original linear problem (\enquote{presmoothing}). Then, the remaining
residual
\begin{equation}\label{eq:mg_res}
  \grd{r}^{(l)} = \grd{\opb}^{(l)} - \opA_l \, \grd{u}^{(l)}
\end{equation}
is restricted to source the linear problem~\eqref{eq:mg_problem} on the
next-coarser grid,
\begin{equation}
  \grd{b}^{(l + 1)} = \restr^{\, l \rightarrow l+1} \, \grd{r}^{(l)}
  \text{.}
\end{equation}
Once presmoothing is complete on the coarsest grid (the \enquote{tip} of the
\vcycle{}), we approximately solve \cref{eq:mg_problem} again
(\enquote{postsmoothing}). The solution of the postsmoothing step is
prolongated to the next-finer grid as a correction,
\begin{equation}
  \grd{u}^{(l)} \quad \leftarrow \quad \grd{u}^{(l)} + \prol^{\, l+1 \rightarrow l} \, \grd{u}^{(l + 1)}
  \text{.}
\end{equation}
Prolongation, correction, and postsmoothing proceed until we have returned to
the finest grid, where the correction and postsmoothing apply to the original
linear problem. Our choice of presmoother and postsmoother to approximately solve
\cref{eq:mg_problem} is detailed in \cref{sec:schwarz} below. Note that on the
coarsest level we apply both presmoothing and postsmoothing.

The restriction of residuals to the next-coarser grid and the prolongation of
corrections to the next-finer grid incur soft synchronization points. Specifically,
only once \emph{all} elements on the finer grid have restricted their residuals
to the coarser grid can \emph{all} elements on the coarser grid proceed, though
individual coarse-grid (parent) elements can already proceed once only their
corresponding fine-grid (child) elements have sent the restricted residuals. The
same applies to the prolongation of corrections from the parent elements back to
their children. Therefore, parent elements on coarser grids ideally follow their
children when distributed among the available cores, initially and at
load-balancing operations.

In contrast to Krylov-subspace algorithms, the multigrid algorithm involves no
global synchronization points. In particular, for diagnostic output we perform a
reduction to compute the global residual norm $\lVert\grd{r}^{(l)}\rVert$ on
every grid, but do so asynchronously in order to avoid a synchronization that is
not algorithmically necessary. Therefore, we do not use the global residual norm
as a convergence criterion for the multigrid solver. Instead, we run a fixed
number of multigrid \vcycle{}s, and typically only a single one to precondition a
Krylov-subspace solver.

\subsection{Schwarz smoother}\label{sec:schwarz}

On every level of the grid hierarchy the multigrid solver relies on a
\emph{smoother} that approximately solves \cref{eq:mg_problem}. In principle,
the smoother can be any linear solver, including a Krylov-subspace solver as
detailed in \cref{sec:lin}. However, to achieve good parallel performance we
have developed a highly asynchronous additive Schwarz
smoother~\cite{Lottes2005-wc, Stiller2016, Vincent2019qpd}, that we employ for
presmoothing and postsmoothing on every level of the multigrid hierarchy. Note that we
also apply it on the coarsest grid, where some authors apply a dedicated
\emph{bottom smoother} instead~\cite{AlOnazi:2017, Kang2015}. Since our coarsest
grids are rarely reduced to a single element (see \cref{sec:mg_hierarchy}), we have, so far, preferred the
asynchronous Schwarz smoother over direct bottom smoothers. Our Schwarz smoother
can be used standalone, as a preconditioner for a Krylov-type linear solver, or
as a smoother for the multigrid solver (which may, in turn, precondition a
Krylov-type solver).

The additive Schwarz method works by solving many subproblems in parallel and
combining their solutions as a weighted sum to converge towards the global
solution. The decomposition into independent subproblems makes this linear
solver very parallelizable. The Schwarz solver is based on our prototype in
\ccite{Vincent2019qpd}, with variations to the subdomain geometry and weighting
to take better advantage of our task-based parallelism, and with novel subdomain
preconditioning techniques.

\subsubsection{Subdomain geometry}\label{sec:subd_geometry}

\begin{figure}
  \centering
  \tikzinput{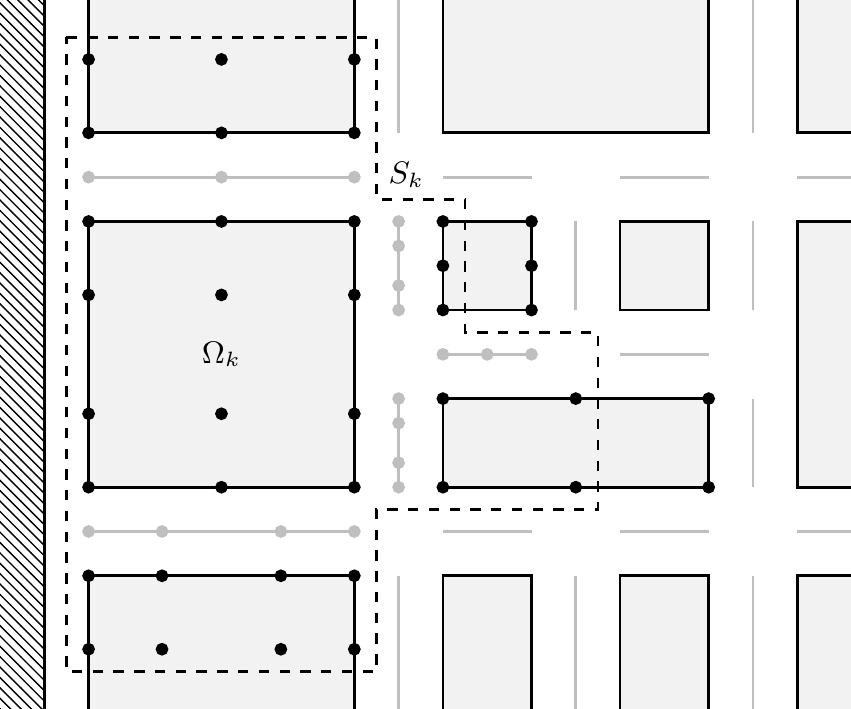}
  \caption{
    \label{fig:subdomain_structure}
    An element-centered subdomain $S_k$ with ${N_\mathrm{overlap} = 2}$ (dashed
    line) associated with the element $\Omega_k$ in a two-dimensional
    computational domain. The domain is composed of elements (black rectangles)
    with their mesh of grid points (black dots) and depicted here in block-logical
    coordinates. The diagonally-shaded region to the left illustrates an
    external domain boundary. The light gray lines between neighboring element
    faces illustrate mortar meshes, which are relevant for the subdomain
    operator in a DG context but play no role in the Schwarz
    algorithm~\cite{dgscheme}. Note that the empty space between the elements in
    this visualization is not part of the computational domain.}
\end{figure}

We partition the computational domain into overlapping, element-centered
subdomains $S_k \subset \Omega$, which have a one-to-one association with the DG
elements $\Omega_k$. Each subdomain $S_k$ is centered on the DG element
$\Omega_k$. It extends by $N_\mathrm{overlap}$ collocation points into
neighboring elements across every face of $\Omega_k$, up to, but excluding, the
collocation points on the face of the neighbor pointing away from the subdomain.
\cref{fig:subdomain_structure} illustrates the geometry of our element-centered
subdomains.

The subdomain does not extend into corner or edge neighbors, which is a choice
different to both \ccite{Stiller2016} and \ccite{Vincent2019qpd}. We avoid
diagonal couplings because in a DG context information only propagates across
faces, as already noted in \ccite{Stiller2016}. Elimination of the corner and
edge neighbors reduces the complexity of the subdomain geometry,
the number of communications necessary to exchange data between elements in the
subdomain, and hence the connectivity of the dependency graph between tasks.
This element-centered subdomain geometry based solely on face neighbors has
proven viable for the test problems presented below, and for our
task-based parallel architecture.

The one-to-one association between elements and subdomains allows to store all
quantities that define the subdomain geometry local to the central element,
i.e., on the same core. The same applies to all data on the grid points of the
subdomain. Therefore, operations local to the subdomain require no
communication, but communication between overlapping elements is necessary to
assemble data on the subdomains, and to make data on subdomains available to
overlapped elements.

\subsubsection{Subdomain restriction}

To restrict quantities defined on the full computational domain $\Omega$ to a
subdomain $S \subset \Omega$ the Schwarz solver employs a \emph{restriction
operator}~$\restr_S$. Since our subdomains are subsets of the grid points in the
full computational domain, our restriction operator simply discards all nodal
data on grid points outside the subdomain. Similarly, the transpose of the
restriction operator, $\restr_S^T$, extends subdomain data with zeros on all
grid points outside the subdomain.\footnote{See also Sec.~3.1 in
\ccite{Stiller2016} for details on the subdomain restriction operation.}

The Schwarz solver also relies on a restriction of the global linear operator
$\opA$ to the subdomains. The subdomain operator $\opA_{S}$ on a subdomain $S$
is formally defined as $\opA_S = \restr_S \opA \restr_S^T$. In practice, it evaluates the
same DG matrix-vector product as the full operator $\opA$, i.e., the left-hand
side of \cref{eq:dg_op_primal}, but assumes that all data outside the subdomain
is zero. It performs all interelement operations of the full DG operator, but
computes them entirely with data local to the subdomain. Therefore, it requires
no communication between cores, as opposed to the global linear operator $\opA$
that must communicate data between nearest neighbors for every
operator application.

\subsubsection{Subdomain problems}

On every subdomain $S$ we solve the restricted problem
\begin{equation}\label{eq:schwarz_subd_problem}
  \opA_S \, \Delta\grd{u}^{(S)} = \grd{r}^{(S)}
\end{equation}
for the subdomain correction $\Delta\grd{u}^{(S)}$. Here, $\opA_S$ is the
subdomain operator and $\grd{r}^{(S)} = \restr_S \, \grd{r}$ is the global
residual $\grd{r} = \grd{\opb} - \opA \grd{u}$ restricted to the subdomain.

The subdomain problems~\eqref{eq:schwarz_subd_problem} are solved by means of a
\emph{subdomain solver}, detailed in \cref{sec:subd}. The choice of subdomain solver
affects only the performance of the Schwarz algorithm, not its convergence or
parallelization properties, assuming the solutions to the subdomain
problems~\eqref{eq:schwarz_subd_problem} are sufficiently precise.

\subsubsection{Weighting}

Once we have obtained the subdomain correction~$\Delta\grd{u}^{(S)}$ on every
subdomain~$S$, we combine them as a weighted sum to correct the solution,
\begin{equation}\label{eq:schwarz_weighting}
  \grd{u} \quad \leftarrow \quad \grd{u} + \sum_S \restr_S^T \left(
    \grd{w}^{(S)} \, \Delta\grd{u}^{(S)}\right)
  \text{,}
\end{equation}
where $\grd{w}^{(S)}$ is a weight at every grid point of the subdomain. This is
the \emph{additive} approach of the algorithm, which has the advantage over
\emph{multiplicative} Schwarz methods that all subdomain problems decouple and
can be solved in parallel.\footnote{See also Sec.~3.1 in \ccite{Stiller2016}
for details on multiplicative variants of Schwarz algorithms.} The weighted sum,
\cref{eq:schwarz_weighting}, is never assembled globally. Instead, every element
adds the contribution from its locally centered subdomain and from all
overlapping subdomains to the components of $\grd{u}$ that resides on the element.

\begin{figure}
  \centering
  \includegraphics[width=\columnwidth]{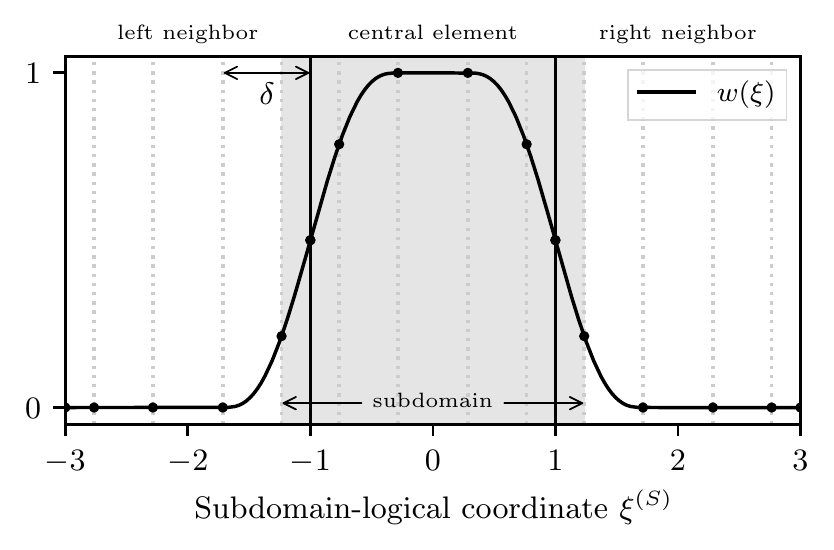}
  \caption{
    \label{fig:schwarz_weight_function}
    The one-dimensional weight function $w(\xi)$ for the Schwarz solver.
    Depicted is an element-centered subdomain in one dimension with
    $N_\text{overlap} = 2$. Every element has $N_k = 6$ LGL collocation points,
    which includes grid points on the shared element boundaries (black vertical
    lines). The overlap width $\delta$ is the logical coordinate distance to the
    first point outside the subdomain, where the weight becomes zero.}
\end{figure}

The weights $\grd{w}^{(S)}$ represent a scalar field on every subdomain, which
must be conserved as
\begin{equation}
  \sum_S \restr_S^T \, \grd{w}^{(S)} = \grd{1}
  \text{.}
\end{equation}
We follow \ccites{Vincent2019qpd,Stiller2016} in constructing the
weights as quintic smoothstep polynomials, but must account for the missing
weight from corner and edge neighbors. Specifically, we compute
\begin{equation}
  w^{(S)}_p = W(\bm{\xi}^{(S)}_p)
\end{equation}
by evaluating the scalar weight function $W(\bm{\xi})$ at the logical
coordinates $\bm{\xi}^{(S)}_p$ of the grid points in the subdomain. These
subdomain-logical coordinates coincide with the element-logical coordinates of
the central element, and extend outside the central element such that $\xi^{(S)}
= \pm 3$ coincides with the sides of the overlapped neighbors that face away
from the subdomain (see abscissa of \cref{fig:schwarz_weight_function}). The
scalar weight function
\begin{equation}\label{eq:schwarz_weight_prod}
  W(\bm{\xi}) = \prod^d_{i=0} w(\xi^i)
\end{equation}
is a product of one-dimensional weight functions,
\begin{align}\label{eq:schwarz_weight_function}
  w(\xi) &= \frac{1}{2}\left( \phi\left( \frac{\xi + 1}{\delta} \right) -
  \phi\left( \frac{\xi - 1}{\delta} \right) \right) \\
  \text{with} \quad
  \phi(\xi) &= \begin{cases}
      \frac{1}{8}\left(15\xi - 10\xi^3 + 3\xi^5\right)
      &\quad \xi \in [-1, 1] \\
      \mathrm{sign}(\xi) \quad
      &\quad |\xi| > 1
      \text{,}
    \end{cases}
\end{align}
where $\phi(\xi)$ is a second-order smoothstep function, i.e., a quintic
polynomial, and $\delta \in (0,2]$ is the \emph{overlap width}. The overlap
width is the logical coordinate distance from the boundary of the central
element to the first collocation point \emph{outside} the overlap region
(see \cref{fig:schwarz_weight_function}). With
this definition the overlap width is nonzero even when the overlap extends only
to a single LGL point in the neighbor, which coincides
with the element boundary. Furthermore, the
weight is always zero at subdomain-logical coordinates $\xi^{(S)}=\pm 3$, even
for $\delta = 2$ when the overlap region covers the full neighbor in width. This
is the reason we never include the collocation points on the side of
the neighbor facing away from the subdomain (see \cref{sec:subd_geometry}).
\Cref{fig:schwarz_weight_function} illustrates the shape of the weight function.

To account for the missing weight from corner and edge neighbors, we could add
it to the central element, to the overlap data from face neighbors, or split it
between the two. We choose to add it to the face neighbors that share a corner
or edge, since in a DG context that is where the information from those regions
propagates through.

\subsubsection{Algorithm}

\begin{figure}
  \centering
  \tikzinput{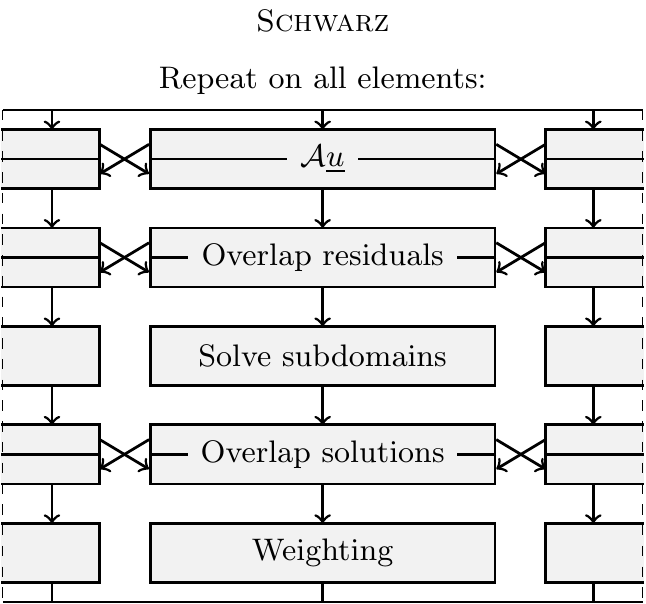}
  \caption{
    \label{fig:schwarz_tasks}
    Parallelization structure of the task-based Schwarz smoother (\cref{sec:schwarz}).}
\end{figure}

\cref{fig:schwarz_tasks} illustrates our task-based implementation of the
additive Schwarz solver. In each iteration, the algorithm computes the residual
$\grd{r}=\grd{b} - \opA \grd{u}$, restricts it to all subdomains as
$\grd{r}^{(S)}=\restr_S \, \grd{r}$, and exchanges it on overlap regions with
neighboring elements. Once an element has received all residual data on its
subdomain, it solves the subdomain problem, \cref{eq:schwarz_subd_problem}, for
the correction~$\Delta \grd{u}^{(S)}$. Since all elements perform such a
subdomain solve, we end up with a subdomain solution~$\Delta \grd{u}^{(S)}$ on
every element-centered subdomain, and the solutions overlap. Therefore, the
algorithm exchanges the subdomain solutions on overlap regions with neighboring
elements and adds them to the solution field~$\grd{u}$ as the weighted sum,
\cref{eq:schwarz_weighting}.

In order to compute the residual~$\grd{r}$ that is restricted to the
subdomains to serve as source for the subdomain solves, we must apply the global
linear operator~$\opA$ to the solution field~$\grd{u}$ once per Schwarz
iteration. This operator application, as well as the steps to communicate the
residuals and the solutions on overlaps, incur soft synchronization points
through nearest-neighbor couplings. However, once the residuals on overlaps are
communicated, all subdomain solves are independent of each other. This
constitutes the main source of parallelization in the elliptic solver.

The subdomain solves not only run in parallel, but also scale with the problem
size. Increasing the number of grid points in the elements ($p$~refinement) makes
the subdomain solves more expensive, but the effectiveness of a Schwarz
iteration ideally remains the same. Increasing the number of elements
($h$~refinement) leads to more subdomain solves that can run in parallel. The
Schwarz solver does not resolve large-scale modes, so Krylov-type solvers still
rely on the multigrid algorithm to scale with $h$~refinement.

\subsection{Subdomain solver}\label{sec:subd}

Once overlapping subdomains have exchanged data we can solve all subdomain
problems~\eqref{eq:schwarz_subd_problem} in parallel with data local to each
subdomain. Since the subdomain operator $\opA_S$ is defined as a matrix-vector
product, we solve \cref{eq:schwarz_subd_problem} with a preconditioned GMRES
algorithm, or with conjugate gradients for symmetric
positive definite problems. The algorithm is the same as detailed in
\cref{sec:lin} for our task-based parallel Krylov-subspace linear solver, but
implemented separately as a serial algorithm. In future work, we may opt to
parallelize the subdomain solver over a few threads with shared memory, but
currently we prefer to employ the available cores to solve multiple subdomain
problems in parallel. In particular, on coarse multigrid levels where the number
of elements can be smaller than the number of available cores, parallelizing
the subdomain solves over otherwise idle cores may increase performance.

The iterative Krylov subdomain solves constitute the majority of the total computational expenses, so a
suitable preconditioner for them can speed up the elliptic solve significantly.
To our knowledge, preconditioners for Schwarz subdomain solvers have gotten
little attention in the literature so far. In some cases, the discretization
scheme allows to construct a matrix representation for the subdomain operator
explicitly, making it possible to invert it directly with little
effort~\cite{Stiller2016}. In other cases, the subdomain operator is small enough
to build the matrix representation column-by-column (see
\cref{sec:explicit_inverse_solver}), e.g., when solving the Poisson equation.
However, when solving sets of coupled elliptic equations the subdomain operator
can easily become too large to construct explicitly. For example, the subdomain
operator for the XCTS equations~\eqref{eq:xcts} (five variables) on a
three-dimensional grid with $8^3$ grid points per element and
$N_\mathrm{overlap}=2$ is a matrix of size
$6400 \times 6400$. Stored densely, it requires over \SI{300}{\mega\byte} of
memory per element, so typical contemporary computing clusters with a few
\si{\giga\byte} of memory per core could only hold a few elements per core.
Sparse storage reduces the memory cost significantly, but still requires \num{6400}
subdomain-operator applications to construct the matrix representation and a
nonnegligible cost to invert and to apply it. With an iterative Krylov-subspace
algorithm and a suitable preconditioner we can solve the subdomain problems on
an element with significantly lower cost and memory requirements. For example,
test problem \ref{sec:kerrschild} completes about an order of magnitude faster
with the subdomain preconditioner laid out in this section, than with an
unpreconditioned GMRES subdomain solver.

\subsubsection{Laplacian-approximation preconditioner}\label{sec:laplacian_approx}

We support the iterative subdomain solver with a
Laplacian-approximation preconditioner. It approximates the linearized
elliptic PDEs with a Poisson equation for every variable.
A similar preconditioning strategy has proven successful for the \spec{}
code~\cite{Pfeiffer2003-mt}, but in the context of a spectral discretization
scheme and a very different linear-solver stack. Specifically, we approximate
the subdomain problem, \cref{eq:schwarz_subd_problem}, as a set of independent
Poisson subdomain problems
\begin{equation}\label{eq:lapl_approx}
  \opA^\mathrm{Poisson}_S \, \Delta\grd{u}^{(S)}_A = \grd{r}^{(S)}_A
  \text{,}
\end{equation}
where the index $A$ iterates over all primal variables (see
\cref{sec:discretization}). Here, $\opA^\mathrm{Poisson}_S$ is the
DG-discretization of the negative Laplacian $-g^{ij}\nabla_i\nabla_j$ on the
subdomain according to \cref{sec:discretization}. For example, a
three-dimensional XCTS problem has five variables, so \cref{eq:lapl_approx}
approximates the linearization~\eqref{eq:linearized} of the five
equations~\eqref{eq:xcts} as
\begin{equation}
  \bar{\nabla}^2 \delta\psi=0 \text{,} \quad
  \bar{\nabla}^2 \delta(\alpha\psi)=0 \quad \text{and} \quad
  \bar{\nabla}^2 \delta\beta^i=0
  \text{.}
\end{equation}

Depending on the elliptic system at hand we either choose a flat background
metric $g_{ij}=\delta_{ij}$, or the background metric of the elliptic system,
such as the conformal metric $g_{ij}=\bar{\gamma}_{ij}$ for an XCTS system. A
curved background metric reduces the sparsity of the Poisson operator but
approximates the elliptic equations better. In practice, we have found little
difference in runtime between the flat-space and curved-space Laplacian
approximations.

We choose homogeneous Dirichlet or Neumann boundary conditions
for~$\opA^\mathrm{Poisson}_S$. For variables and element faces where the
original boundary conditions are of Dirichlet type we choose homogeneous
Dirichlet boundary conditions, and for those where the original boundary
conditions are of Neumann type we choose homogeneous Neumann boundary
conditions. This may lead to more than one distinct Poisson
operator on subdomains with external boundaries, one per unique combination of
element face and boundary-condition type among the variables. Subdomains
that have exclusively internal boundaries only ever have a single Poisson
operator, which applies to all variables. Note that the choice of
homogeneous boundary conditions for the Poisson subdomain problems is compatible
with inhomogeneous boundary value
problems, because the inhomogeneity in the boundary conditions is absorbed in
the fixed sources when the equations are linearized~\cite{dgscheme}.

To solve the Poisson subdomain problems~\eqref{eq:lapl_approx}, one per
variable, we can (again) employ any choice of linear solver, such as a
(preconditioned) Krylov-subspace algorithm.\footnote{The absurdity of adding
a \emph{third} layer of nested preconditioned linear solvers was not lost on the
authors.} However, at this point we have reduced the full elliptic problem down
to a single Poisson problem limited to a subdomain that is solved for all
variables, or a few Poisson problems on subdomains with external boundaries.
Therefore, it becomes feasible, and indeed worthwhile, to construct the Poisson
subdomain-operator matrix explicitly and to invert it directly. In particular, the
approximate Poisson subdomain-operator matrix remains valid throughout the full
nonlinear elliptic solve, as long as the grid, the background metric, and the
type of boundary conditions remain unchanged, so that its construction cost is
amortized over many applications.

\subsubsection{Explicit-inverse solver}\label{sec:explicit_inverse_solver}

We solve the Poisson subproblems of the Laplacian-approximation preconditioner,
\cref{eq:lapl_approx}, with an explicit-inverse solver.
It constructs the matrix representation of a linear
subdomain operator~$\opA_S$ column-by-column, and then inverts it directly by
means of an LU decomposition. Once the inverse~$\opA_S^{-1}$ has been
constructed, each subdomain problem $\opA_S\,\Delta\grd{u}^{(S)} = \grd{r}^{(S)}$ is solved by
a single application of the inverse matrix,
\begin{equation}
  \Delta\grd{u}^{(S)} = \opA_S^{-1} \, \grd{r}^{(S)}
  \text{.}
\end{equation}
This means that subdomains have a large initialization cost, but fast repeated
solves.

When the explicit-inverse solver is employed as a preconditioner, e.g., to solve
the individual Poisson problems of the Laplacian-approximation preconditioner
(\cref{sec:laplacian_approx}), the inverse does not need to be exact. Therefore,
we construct an \emph{incomplete} LU decomposition with a configurable fill-in
and store it in sparse format. Then, each subdomain problem reduces to two
sparse triangular matrix solves. We use the \texttt{Eigen}~\cite{eigenweb}
sparse linear algebra library for the incomplete LU decomposition, which uses the
\texttt{ILUT} algorithm~\cite{Saad1994}. The Poisson subdomain-operator
matrices~$\opA_S^\mathrm{Poisson}$
have a sparsity of about \SI{90}{\percent}, which translates to a sparsity of
about \SI{90}{\percent} for the incomplete LU decomposition as well, since we
use a fill-in factor of one. The sparsity of the inverse reduces the
computational cost for applying it to every subdomain problem, as well as the
memory required to store the inverse.

Note that the explicit matrix must be reconstructed when the linear operator
changes. The Poisson operators of the Laplacian-approximation preconditioner do
not typically change, which makes the explicit-inverse solver very effective
(see \cref{sec:laplacian_approx}). However, in case we apply the
explicit-inverse solver to the full subdomain problem directly, the linearized
operator typically changes between every outer nonlinear solver iteration.
In such cases, we can choose to skip the
reconstruction of the explicit matrix to avoid the computational expense, at
the cost of losing accuracy of the solver. When the reconstruction is skipped,
the cached matrix only approximates the subdomain operator, but can still provide
effective preconditioning.

\section{Test problems}\label{sec:tests}

\begin{figure*}
  \centering
  \subfloat[Initial error $\grd{u}_0 - \grd{u}_\mathrm{analytic}$]{
    \includegraphics[width=0.4\textwidth,trim=150px 300px 150px 200px,clip]{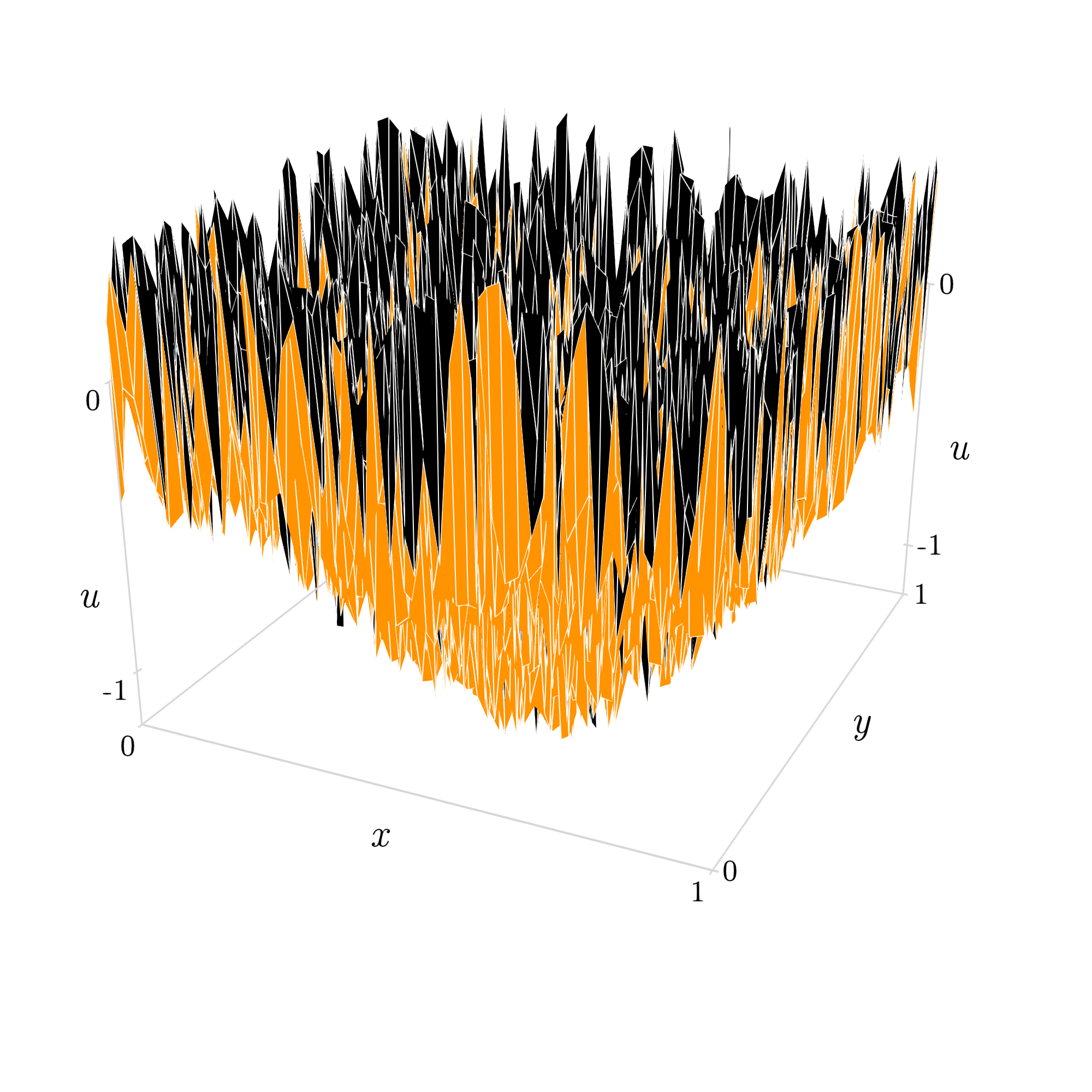}
    \label{fig:poisson_smoothing_initial}
  }\hfil
  \subfloat[Error after 6 unpreconditioned GMRES iterations]{
    \includegraphics[width=0.4\textwidth,trim=150px 300px 150px 200px,clip]{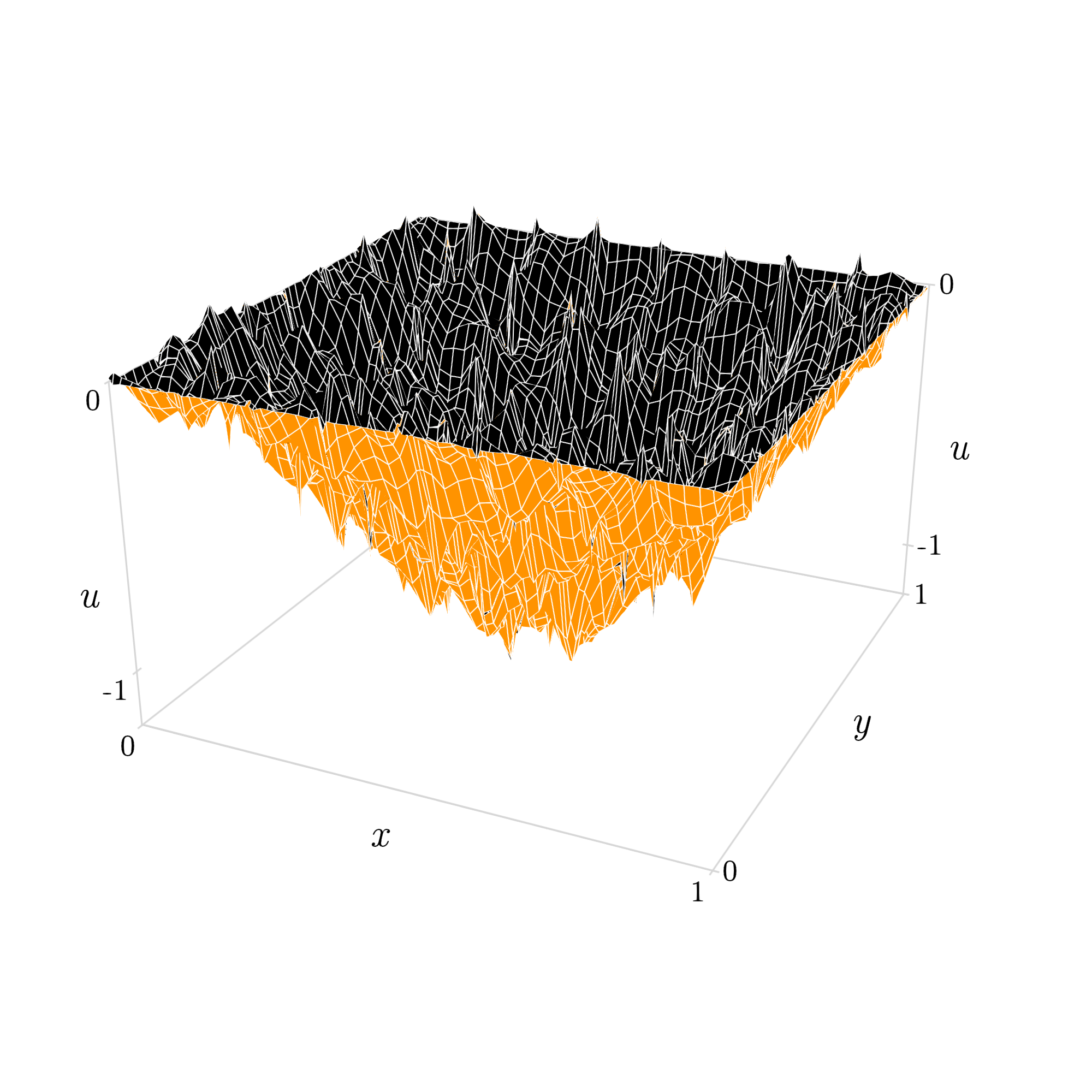}
    \label{fig:poisson_smoothing_noprec}
  }\\
  \subfloat[Error after 6 Schwarz-smoothing iterations]{
    \includegraphics[width=0.4\textwidth,trim=150px 300px 150px 200px,clip]{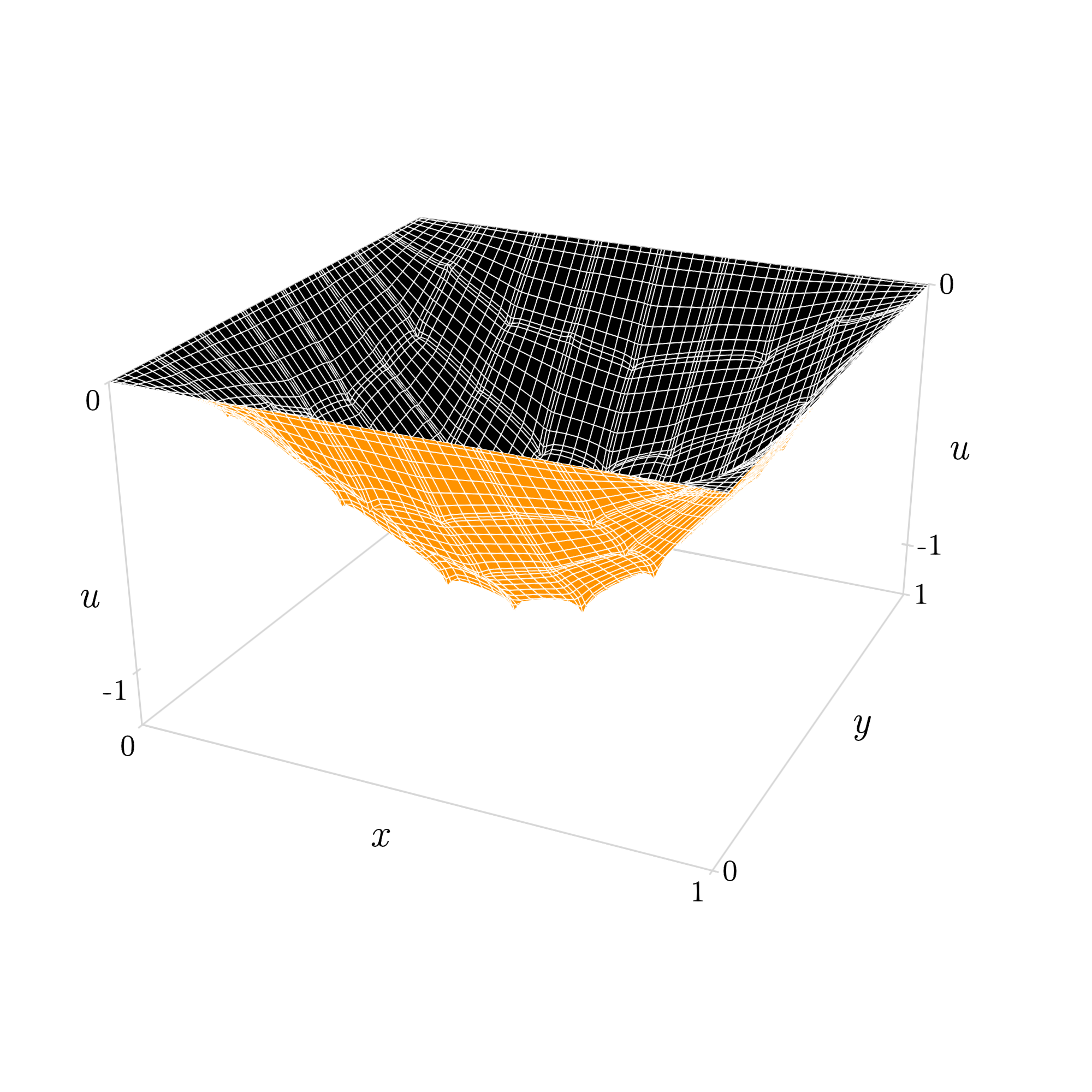}
    \label{fig:poisson_smoothing_schwarz}
  }\hfil
  \subfloat[Error after 1 multigrid-Schwarz \vcycle{}]{
    \includegraphics[width=0.4\textwidth,trim=150px 300px 150px 200px,clip]{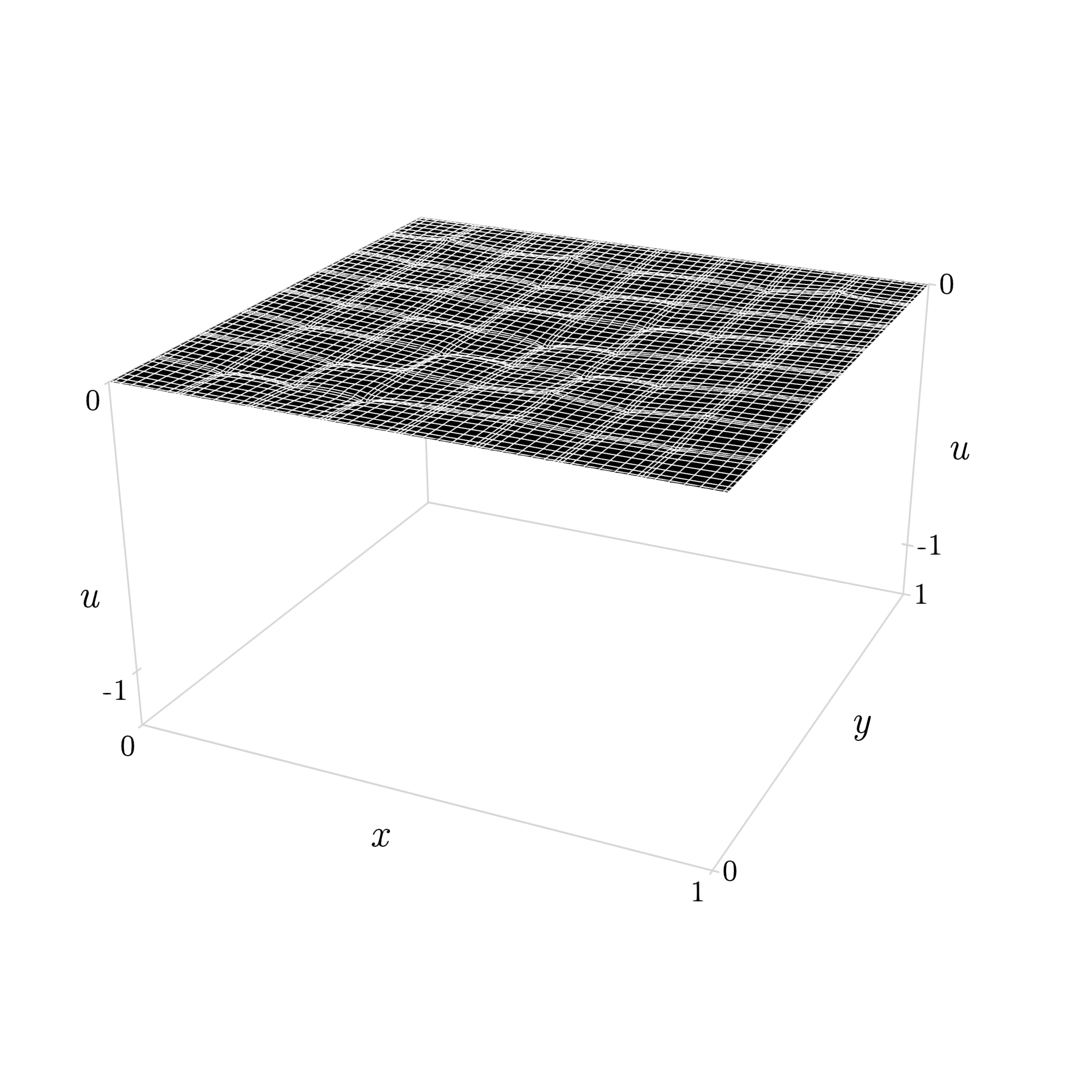}
    \label{fig:poisson_smoothing_mg}
  }
  \caption{
    \label{fig:poisson_smoothing}
    Intermediate errors of the 2D sinusoidal Poisson problem
    (\cref{sec:poisson}) with different components of the elliptic solver.
    Panel~\protect\subref{fig:poisson_smoothing_initial} shows the error of the
    random initial guess.
    Panels~\protect\subref{fig:poisson_smoothing_noprec}--\protect\subref{fig:poisson_smoothing_mg}
    show the error after six applications of the linear operator.}
\end{figure*}

The following numerical tests demonstrate the accuracy, scalability, and parallel
efficiency of the elliptic solver on a variety of linear and nonlinear elliptic
problems.

All computations were performed on our local computing cluster \texttt{Minerva}.
It is composed of 16-core nodes, each with two eight-core Intel Haswell
E5-2630v3 processors clocked at \SI{2.40}{\giga\hertz} and \SI{64}{\giga\byte}
of memory, connected with an Intel Omni-Path network. We distribute elements
evenly among cores following the strategy detailed in \cref{sec:algs}, leaving
one core per node free to perform communications.

\subsection{A Poisson problem}\label{sec:poisson}

\begin{figure}
  \centering
  \includegraphics[width=\columnwidth]{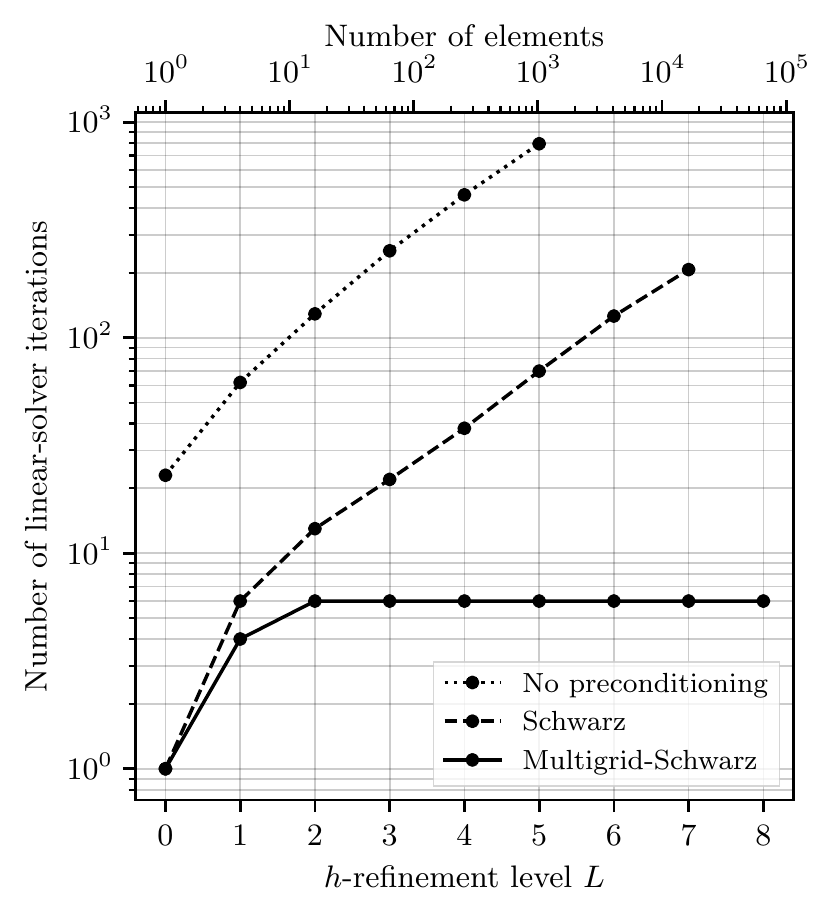}
  \caption{
    \label{fig:poisson_iterations}
    Number of linear-solver iterations for the Poisson problem
    (\cref{sec:poisson}). The multigrid-Schwarz preconditioner achieves
    scale independence.}
\end{figure}

As a first test we solve the flat-space Poisson equation in two dimensions,
\begin{equation}\label{eq:poisson}
  -\partial_i\partial_i u(\bm{x}) = f(\bm{x})
  \text{,}
\end{equation}
for the solution
\begin{equation}\label{eq:prodofsin}
  u_\text{analytic}(\bm{x})=\sin{(\pi x)}\sin{(\pi y)}
\end{equation}
on a rectilinear domain $\Omega=[0,1]^2$ with Dirichlet boundary conditions.
This problem is also studied
in the context of multigrid-Schwarz methods, with slight variations, in
\ccites{Lottes2005-wc, Stiller2016}. To obtain the solution~\eqref{eq:prodofsin}
numerically we choose the fixed source~$\dgf(\bm{x})=2\pi^2\sin{(\pi
x)}\sin{(\pi y)}$, select homogeneous Dirichlet boundary conditions
${u^\mathrm{b}=0}$, and solve the DG-discretized problem~\eqref{eq:dg_residuals}
with penalty parameter~${C=1}$. We have evaluated the properties of the
DG-discretized operator
for this problem in \ccite{dgscheme}. To assess the convergence behavior of the
elliptic solver for this test problem we choose an initial guess $\grd{u}_0$,
where each value is uniformly sampled from $[-0.5,0.5]$.

\Cref{fig:poisson_smoothing} illustrates the effectiveness of our
algorithm in resolving small-scale and large-scale modes in the solution.
Plotted is the error to the analytic solution, $\grd{u} - \grd{u}_\mathrm{analytic}$.
\Cref{fig:poisson_smoothing_initial} depicts the initial error on a
computational domain that is partitioned into~$8 \times 8$ quadratic elements
with~$9 \times 9$ grid points each.
\Crefrange{fig:poisson_smoothing_noprec}{fig:poisson_smoothing_mg} present the
error after six applications of the linear operator, but with different
components of the elliptic solver enabled. \Cref{fig:poisson_smoothing_noprec}
employs no preconditioning at all, thus reaches six operator applications after
six iterations of the GMRES algorithm. It resolves some of the random
fluctuations, but retains the large-scale sinusoidal error.
\Cref{fig:poisson_smoothing_schwarz} preconditions every GMRES iteration with
six Schwarz-smoothing steps, thus reaches six operator applications after a
single GMRES iteration. The Schwarz smoother uses $N_\mathrm{overlap} = 2$. It
resolves most of the random fluctuations, but retains the large-scale error.
Note that the six operator applications do not accurately reflect the
computational expense to arrive at \cref{fig:poisson_smoothing_schwarz}, because
a significant amount of work is spent on the subdomain solves. We employ the
explicit-inverse subdomain solver directly here to solve subdomain problems
because the Laplacian-approximation preconditioner is redundant for a pure
Poisson problem (see \cref{sec:subd}), but note that the subdomain solver has no
effect on the results depicted in \cref{fig:poisson_smoothing} as long as it is
sufficiently precise. Finally, \cref{fig:poisson_smoothing_mg} preconditions
every GMRES iteration with a single four-level multigrid-Schwarz \vcycle{}. The
\vcycle{} employs three Schwarz presmoothing and postsmoothing steps on every level, thus
reaches six operator applications on the finest grid after a single GMRES
iteration. Again, the number of operator applications is not entirely
representative of the computational expense because it disregards the work done
on coarser levels. The \vcycle{} successfully resolves the large-scale error.

\Cref{fig:poisson_iterations} presents the number of GMRES iterations that the
elliptic solver needs to reduce the magnitude of the residual by a factor
of~$10^{10}$, for a series of $h$-refined domains. We construct $h$-refinement
levels~$L$ by repeatedly splitting all elements in the rectangular domain in two
along both dimensions. All elements have~$6 \times 6$ grid points.
Shown in \cref{fig:poisson_iterations} are an unpreconditioned GMRES algorithm,
a GMRES algorithm preconditioned with three Schwarz-smoothing steps per
iteration, and a GMRES algorithm preconditioned with one multigrid-Schwarz
\vcycle{} per iteration.
The number of multigrid levels is equal to the number $L$ of refinement levels,
so that the coarsest level always covers the entire domain with a single element.
Every level runs three presmoothing and postsmoothing
steps, and subdomains have $N_\mathrm{overlap} = 2$. The Schwarz preconditioning
alone reduces the number of iterations by a factor of $\sim 10$, but does not
affect the scaling with element size. However, the multigrid-Schwarz
preconditioning removes the scaling entirely, meaning the number of GMRES
iterations remains constant even when the domain is partitioned into more and
smaller elements. The multigrid algorithm achieves this scale independence
because it supports the iterative solve with information from coarser grids,
including large-scale modes in the solution that span the entire domain.
Each preconditioned iteration is typically more computationally expensive than
an unpreconditioned iteration, but the preconditioner reduces the number of
iterations such that the solve completes faster overall.\footnote{Note that the cost of
unpreconditioned GMRES iterations is eventually dominated by the
orthogonalization procedure (see \cref{sec:lin}), which slows down the
unpreconditioned solve significantly at large iteration counts. This
effect can be remedied by \emph{restarting} GMRES variants, but at the cost of
possible stagnation. See Sec.~6.5.5 in \ccite{Saad2003} for a discussion.
Conjugate gradient algorithms avoid this issue for symmetric linear operators.}
We find that even for the simple Poisson problem the unpreconditioned algorithm
becomes prohibitively slow around $\mathcal{O}(10^3)$ elements ($L=5$),
approaching an hour of runtime and the memory capacity of our ten compute nodes.
In contrast, the Schwarz preconditioner reduces the runtime to solve the same
problem to below one minute, and the multigrid-Schwarz preconditioner reduces
the runtime to only three seconds.
Crucial to achieving these runtimes at high resolution are the parallelization
properties of the algorithms. In particular, the additional computational
expense that the preconditioner spends Schwarz-smoothing all multigrid levels is
parallelizable within each level. The following test problem
\ref{sec:kerrschild} explores the parallelization in greater detail.

\begin{figure}
  \centering
  \includegraphics[width=\columnwidth]{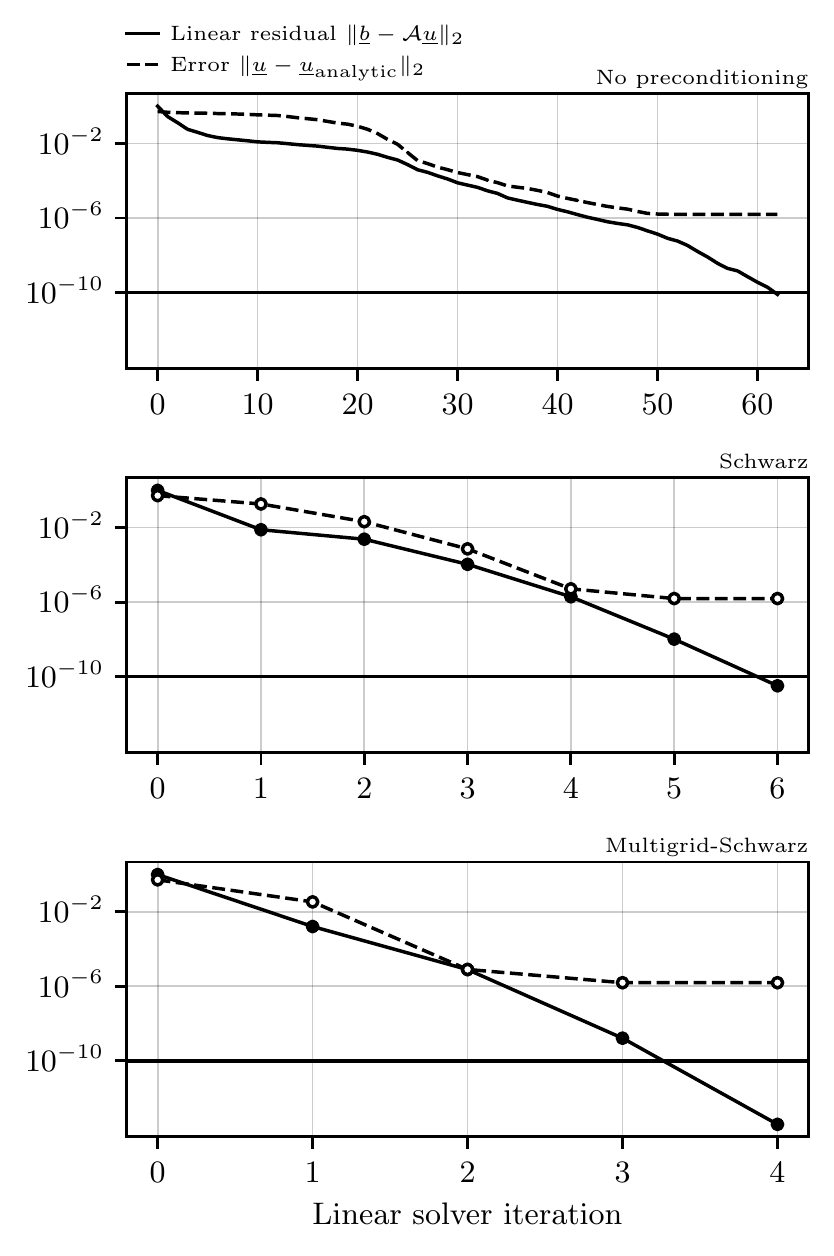}
  \caption{
    \label{fig:poisson_convergence}
    Convergence of the elliptic solver for the linear Poisson problem with
    $h$-refinement level $L=1$. The solid line shows the relative linear-solver
    residual magnitude $\lVert\grd{b} - \opA\grd{u}\rVert_2$, and the dashed
    line shows the error to the analytic solution, $\lVert\grd{u} -
    \grd{u}_\mathrm{analytic}\rVert_2$ as a root mean square over grid points,
    which approaches the DG discretization error.}
\end{figure}

\Cref{fig:poisson_convergence} gives a detailed insight into the convergence
behavior of the elliptic solver for the $L=1$ configuration. Presented is both
the linear-solver residual magnitude $\lVert\grd{b} - \opA\grd{u}\rVert_2$, and
the error to the analytic solution, $\lVert\grd{u} -
\grd{u}_\mathrm{analytic}\rVert_2$. The linear-solver residual (solid line) is
being reduced by a factor of $10^{10}$ by the GMRES algorithm, equipped with the
three different preconditioning configurations explored in
\cref{fig:poisson_iterations}. With no preconditioning, the convergence
stagnates until large-scale modes in the solution are resolved (see also
\cref{fig:poisson_smoothing}). The Schwarz preconditioner reduces the number of
iterations by about an order of magnitude, and the multigrid-Schwarz
preconditioner achieves clean exponential convergence. The error to the analytic
solution (dashed line) follows the convergence of the residual magnitude. Once
the discrete problem $\opA\grd{u} = \grd{b}$, \cref{eq:matrix_eq}, is solved to
sufficient precision, the remaining error $\grd{u} - \grd{u}_\mathrm{analytic}$
is the DG discretization error. It is independent of the computational technique
used to solve the discrete problem, and determined entirely by the
discretization scheme on the computational grid, as summarized in
\cref{sec:discretization} and detailed in \ccite{dgscheme}.\footnote{For a study
of the DG discretization error for this problem see Fig.~7 in \ccite{dgscheme},
where the configuration solved in \cref{fig:poisson_convergence} is circled.}

\subsection{A black hole in general relativity}\label{sec:kerrschild}

Next, we solve a general-relativistic problem involving a black hole.
Specifically, we solve the Einstein constraint equations in the XCTS
formulation, \cref{eq:xcts}, for a Schwarzschild black hole in Kerr-Schild
coordinates. To this end we set the conformal metric and the trace of
the extrinsic curvature to their respective Kerr-Schild quantities,
\begin{subequations}\label{eq:kerrschild_bg}
\begin{align}
  \bar{\gamma}_{ij} &= \delta_{ij} + \frac{2M}{r} l_i l_j
\intertext{and}
  K &= \frac{2M\alpha^3}{r^2}\left(1 + \frac{3M}{r}\right)
  \text{,}
\end{align}
\end{subequations}
where $M$ is the mass parameter, $r=\sqrt{x^2+y^2+z^2}$ is the Euclidean
coordinate distance, and $l^i=l_i=x^i/r$.\footnote{See Table 2.1
in~\ccite{BaumgarteShapiro}.} The time-derivative quantities $\bar{u}_{ij}$ and
$\partial_t K$ in the XCTS equations~\eqref{eq:xcts} vanish, as do the matter
sources $\rho$, $S$, and~$S^i$. With these background quantities specified, the
solution to the XCTS equations is
\begin{subequations}
\begin{align}
  \label{eq:kerrschild_psi}
  \psi &= 1
  \text{,} \\
  \label{eq:kerrschild_lapse}
  \alpha &= \left(1 + \frac{2M}{r}\right)^{-1/2}
  \text{,} \\
  \label{eq:kerrschild_shift}
  \beta^i &= \frac{2M}{r} \alpha^2 l^i
  \text{.}
\end{align}
\end{subequations}
Note that we have chosen a conformal decomposition with $\psi=1$ here, but other
choices of $\psi$ and $\bar{\gamma}_{ij}$ that keep the spatial metric
$\gamma_{ij} = \psi^4 \bar{\gamma}_{ij}$ invariant are equally admissable.

We solve the XCTS equations numerically for the conformal factor~$\psi$, the
product~$\alpha\psi$, and the shift~$\beta^i$. The conformal
metric~$\bar{\gamma}_{ij}$ and the trace of the extrinsic curvature, $K$, are
background quantities that remain fixed throughout the
solve. Note that for this test problem the conformal metric $\bar{\gamma}_{ij}$
is not flat, resulting in a problem formulated on a curved manifold.

\begin{figure}
  \centering
  \includegraphics[width=0.9\columnwidth]{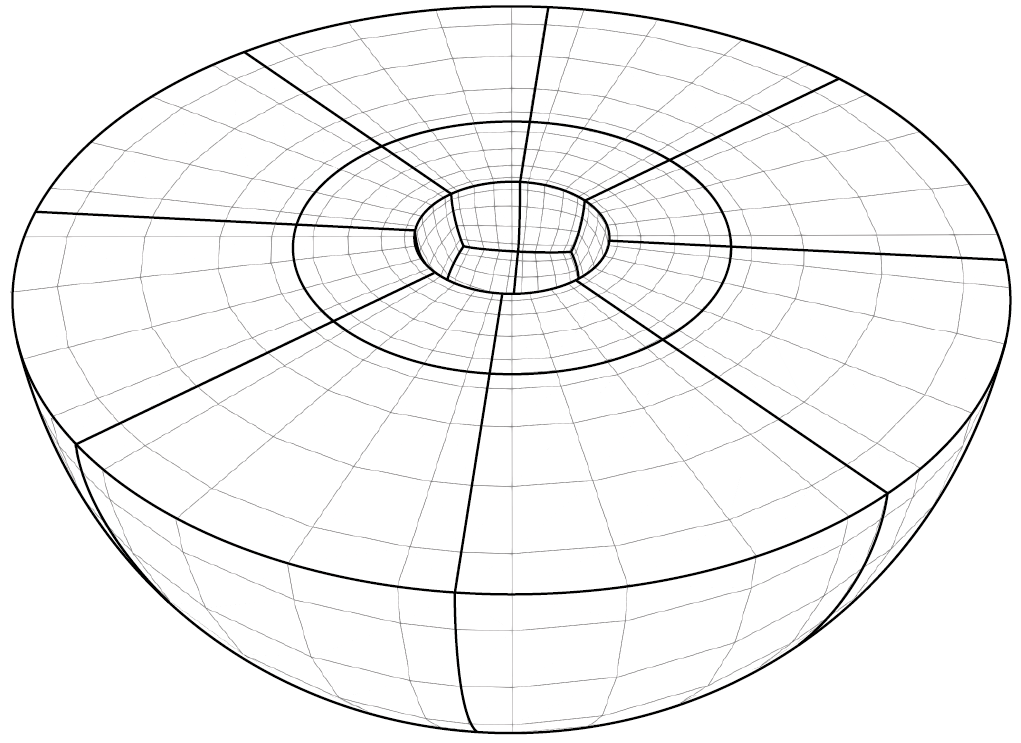}
  \caption{
  \label{fig:kerrschild_domain}
    A cut through the uniformly-refined spherical-shell domain used in the
    black hole problem (\cref{sec:kerrschild}). The domain consists of six
    wedges with a logarithmic radial coordinate map enveloping an excised
    sphere. In this example each wedge is isotropically $h$-refined once, i.e.,
    split once in all three dimensions, resulting in a total of 48 elements.
    Note the elements are split in half along their logical axes, so the element
    size scales logarithmically in radial direction just like the distribution
    of grid points within the elements. Each element has six grid point per
    dimension, so fields are represented as polynomials of degree five.}
\end{figure}

We employ the DG scheme~\eqref{eq:dg_residuals} with penalty parameter~${C=1}$
to discretize the
XCTS equations~\eqref{eq:xcts} on a three-dimensional spherical-shell domain, as
illustrated in \cref{fig:kerrschild_domain}. The domain envelops an excised
sphere that represents the black hole, so it has an inner and an outer external
boundary that require boundary conditions. To obtain the Schwarzschild solution
in Kerr-Schild coordinates we impose
\crefrange{eq:kerrschild_psi}{eq:kerrschild_shift} as Dirichlet conditions at
the outer boundary of the spherical shell at~$r=10M$. We place the inner radius
of the spherical shell at $r=2M$ and impose mixed Dirichlet and Neumann
conditions at the inner boundary. Specifically, we impose the Neumann condition
$n^i\partial_i\psi = 0$ on the conformal factor, and
\crefrange{eq:kerrschild_lapse}{eq:kerrschild_shift} as Dirichlet conditions on
the lapse and shift. The reason for this choice is to mimic apparent-horizon
boundary conditions, as employed in the following test problem (\cref{sec:bbh}).
Choosing apparent-horizon boundary conditions for the Kerr-Schild problem is
also possible, but requires either an initial guess close to the solution to
converge, or a conformal decomposition different from $\psi=1$. The reason is
the strong nonlinearity in the apparent-horizon boundary conditions that takes
the solution away from $\psi_0=1$ initially. We have confirmed this behavior of
the XCTS equations with the \spec{} code, and have presented the convergence
of the DG discretization error with apparent-horizon boundary conditions for
the Kerr-Schild problem in \ccite{dgscheme}. With the simpler Dirichlet and
Neumann boundary condition we can seed the elliptic solver with a flat initial
guess, i.e., $\psi_0=1$, $\alpha_0=1$ and $\beta_0^i=0$, which allows for better
control of the test problem.

\begin{figure}
  \centering
  \includegraphics[width=\columnwidth]{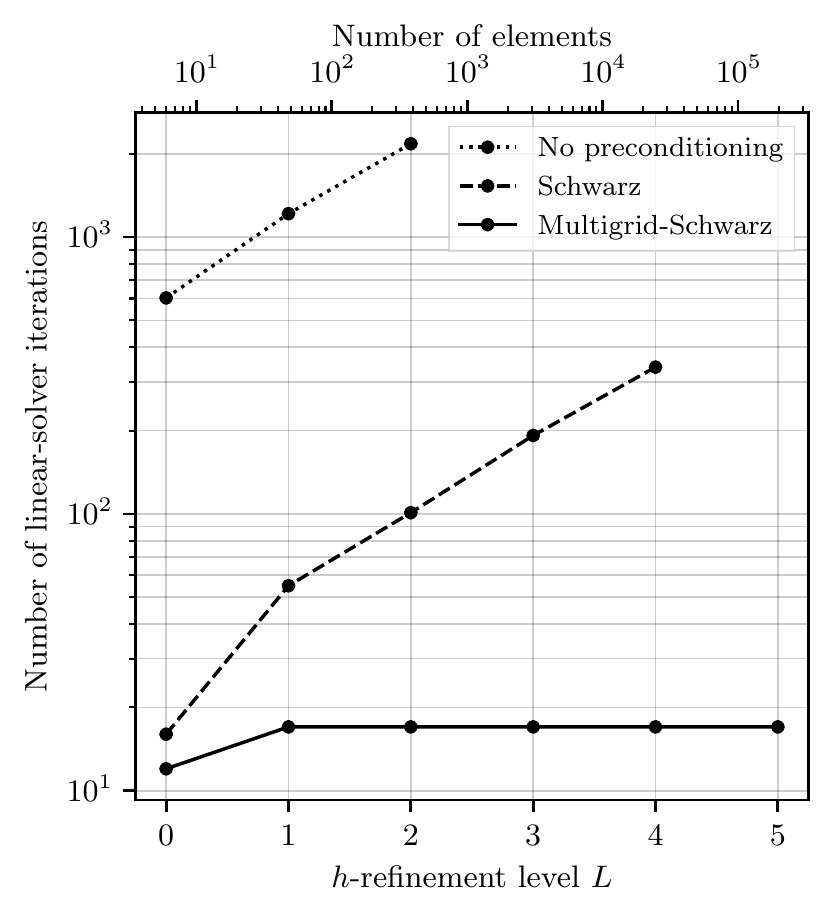}
  \caption{
    \label{fig:kerrschild_iterations}
    Number of linear-solver iterations for the black hole problem
    (\cref{sec:kerrschild}). The multigrid-Schwarz preconditioner achieves
    scale independence. The $L=1$ configuration (48 elements) is pictured in
    \cref{fig:kerrschild_domain}.}
\end{figure}

\begin{figure}
  \centering
  \includegraphics[width=\columnwidth]{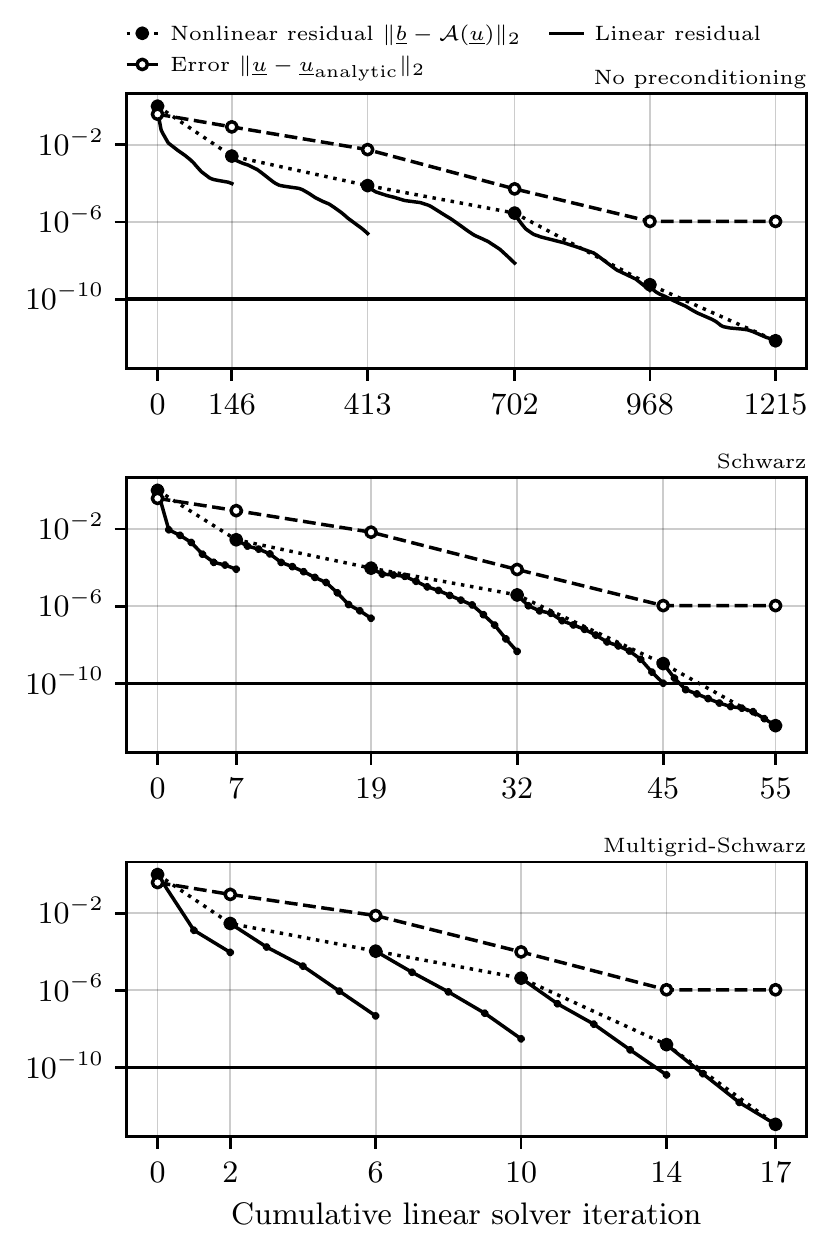}
  \caption{
    \label{fig:kerrschild_convergence}
    Convergence of the Newton-Krylov elliptic solver for the black hole problem
    with $h$-refinement level $L=1$. The dotted line shows the relative residual
    magnitude of the nonlinear solver, $\lVert\grd{b} - \opA\grd{u}\rVert_2$,
    which is driven by a linear solve in every iteration (solid lines, see
    \cref{eq:linearized}). The dashed line shows the error to the analytic
    solution, $\lVert\grd{u} - \grd{u}_\mathrm{analytic}\rVert_2$ as a root mean
    square over all five variables of the XCTS equations,
    $\{\psi,\alpha\psi,\beta^i\}$, and over grid points. It approaches the DG
    discretization error.}
\end{figure}

To assess the convergence behavior of the elliptic solver for this problem we
successively $h$-refine the wedges of the spherical-shell domain into more and
smaller elements, each with six grid points per dimension. We iterate the
Newton-Raphson algorithm until the magnitude of the nonlinear residual has decreased
by a factor of $10^{10}$. In all configurations we have tested, the nonlinear
solver needs five steps and no line-search globalization to reach the target residual. The
linear solver is configured to solve the linearized problem,
\cref{eq:linearized}, by reducing its residual magnitude by a factor of $10^4$.
Schwarz subdomains have $N_\mathrm{overlap} = 2$, and we run three Schwarz presmoothing
and postsmoothing iterations on every multigrid level, including the coarsest.
\Cref{fig:kerrschild_iterations} presents the total number of linear-solver
iterations accumulated over the five nonlinear solver steps. Just like we found
for the simple Poisson problem in \cref{fig:poisson_iterations}, the
multigrid-Schwarz preconditioner achieves scale-independent iteration counts
under $h$~refinement.

\Cref{fig:kerrschild_convergence} presents the convergence behavior of the
elliptic solver for the $L=1$ configuration (pictured in
\cref{fig:kerrschild_domain}) in detail. The convergence of the nonlinear
residual magnitude (dotted line) is independent of the preconditioner chosen for
the linear solver in each iteration (solid lines), since the linearized problems
are solved to sufficient accuracy ($10^{-4}$). Similar to the Poisson problem in
\cref{fig:poisson_convergence}, the multigrid-Schwarz preconditioning achieves
clean exponential convergence, reducing the linear residual by an order of
magnitude per iteration. The nonlinear residual magnitude decreases slowly at
first, when the fields $\grd{u}$ are still far from the solution, and begins to
converge quadratically, following the linear-solver residual, once the fields
are closer to the solution and hence the linearization is more accurate (see
\cref{sec:nonlin}).
The error to the analytic solution (dashed line) approaches the DG
discretization error, as detailed in \cref{sec:poisson}.\footnote{For a study of
the DG discretization error for this problem see Fig.~11 in \ccite{dgscheme},
where the configuration solved in \cref{fig:kerrschild_convergence} is circled.}

\begin{figure}
  \centering
  \includegraphics[width=\columnwidth]{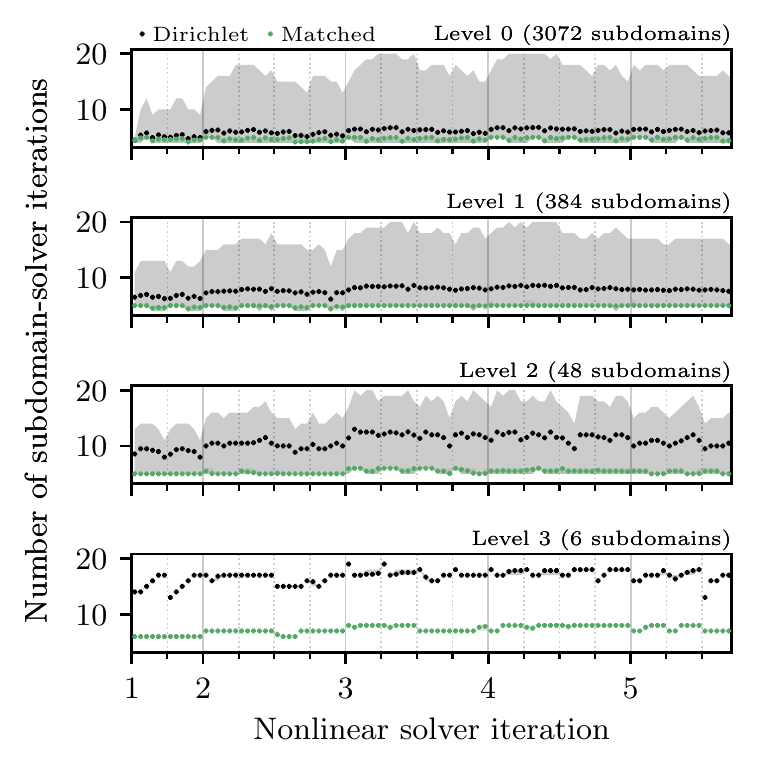}
  \caption{
    \label{fig:kerrschild_subdomain_solves}
    Number of subdomain-solver iterations for the black hole problem
    (\cref{sec:kerrschild}) with the Laplacian-approximation preconditioner. The
    domain is isotropically $h$-refined thrice, so the solve involves four
    multigrid levels. Dots illustrate the average across all subdomains on the
    level, and shaded regions the smallest and largest number of iterations.
    When approximating all equations with a Dirichlet-Laplacian (black),
    subdomains facing the inner excision boundary (see
    \cref{fig:kerrschild_domain}) require more iterations than the average.
    Matching the Laplacian boundary conditions to the problem (green)
    reduces the iteration count and resolves the load imbalance.}
\end{figure}

To solve subdomain problems here we equip the GMRES subdomain solver with the
Laplacian-approximation preconditioner, and solve the five Poisson subproblems
on every subdomain
with the incomplete LU explicit-inverse solver (see \cref{sec:subd}).
\Cref{fig:kerrschild_subdomain_solves} illustrates the importance of matching
the boundary conditions of the approximate Laplacian to the original problem.
When we approximate all five tensor components of the original XCTS problem with
a Dirichlet-Laplacian, ignoring that we impose Neumann-type boundary condition
on $\psi$ at the inner boundary, some subdomains require a significantly larger
number of subdomain-solver iterations than others. We have confirmed that these
subdomains face the inner boundary of the spherical shell. When we use a
Laplacian approximation with matching boundary-condition type for these
subdomains, they need no more subdomain-solver iterations than interior
subdomains.
Specifically, the subdomain preconditioner constructs a Poisson operator matrix
with homogeneous Neumann boundary conditions to apply to the conformal-factor
component of the equations, and another with homogeneous Dirichlet boundary
conditions to apply to the remaining four tensor components. Therefore,
subdomains that face the inner boundary of the spherical shell domain build and
cache two inverse matrices, and all other subdomains build and cache a single
inverse matrix, in the form of an incomplete LU decomposition. Furthermore, when
the Laplacian-approximation preconditioner takes the type of boundary conditions
into account, we find that it is sufficiently precise so we can limit the number
of subdomain-solver iterations to a fixed number. This further balances the load
between elements, decreasing runtime significantly in our tests. Therefore, in
the following we always limit the number of subdomain-solver iterations to
three. With this strategy we find a reduction in runtime of about
\SI{50}{\percent} compared to the naive Dirichlet-Laplacian approximation.

\begin{figure}
  \centering
  \includegraphics[width=\columnwidth]{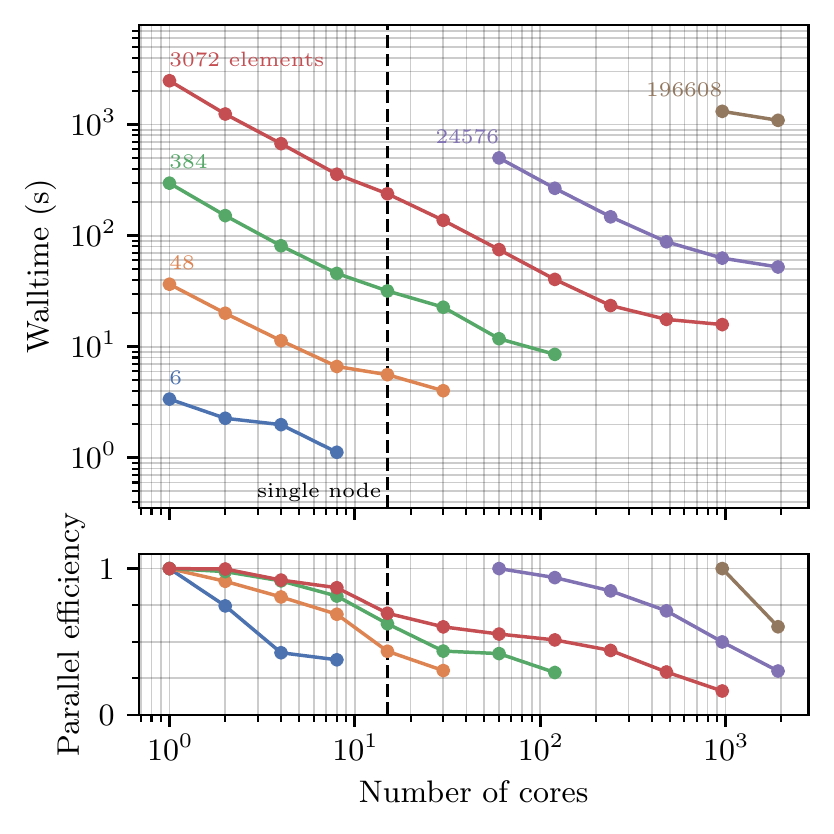}
  \caption{
    \label{fig:kerrschild_parallel}
    Parallel scaling of the black hole problem (\cref{sec:kerrschild}).}
\end{figure}

\Cref{fig:kerrschild_parallel} presents the wall time and parallel efficiency of
the elliptic solves for the black hole problem on up to 2048 cores, which
approaches the capacity of our local computing cluster. We split the domain into
more and smaller elements, keeping the number of grid points in each element
constant at six per dimension, and solve each configuration on a variable number
of cores. These configurations are increasingly expensive to solve, involving up
to 42 million grid points, or \num{\sim 200} million degrees of freedom. They all
complete in at most a few minutes of wall time by scaling up to a few thousand
cores, until they reach the capacity of our cluster. We compute their parallel
efficiency as
\begin{equation}
  \text{Parallel efficiency} = \frac{t_\text{serial}}{t_\text{CPU}}
  \text{,}
\end{equation}
where $t_\text{CPU}=N_\text{cores} \, t_\text{wall}$ is the CPU time of a run,
i.e., the product of wall time and the number of cores, and $t_\text{serial}$ is
the wall time of the configuration runing on a single core. Since configurations
with 24576 elements and more did not complete on a single core in the allotted
time of two hours, we approximate $t_\text{serial}$ with the CPU time of the run
with the lowest number of cores for these configurations, meaning that they
begin at a fiducial parallel efficiency of one. The parallel efficiency
decreases when the number of elements per core becomes small and falls below
\SI{25}{\percent} once each core holds only a few elements.

\Cref{fig:kerrschild_parallel} also shows that the parallel efficiency decreases
more steeply when filling up a single node, than it does when we begin to
allocate multiple nodes. We take this behavior as an indication that shared
hardware resources on a node currently limit our parallel efficiency, which is
an issue also found in \ccite{AlOnazi:2017}. We have confirmed this hypothesis
by running a selection of configurations on the same number of cores, but
distributed over more nodes, so each node is only partially subscribed, and
found that runs speed up significantly. We intend to address this issue in
future optimizations of the elliptic solver. Possible resolutions include better
use of CPU caches, e.g., through a contiguous layout of data on subdomains, or a
shared-memory OpenMP parallelization of subdomain solves, so the cores of a node
are working on a smaller amount of data at any given time.
The parallel efficiency also decreases once we reach the capacity of our
cluster, at which point we expect that communications spanning the full cluster
dominate the computational expense. We intend to test the parallel scaling on
larger clusters with more cores per node in the future. We also plan to
investigate the effect of hyperthreading on the parallel efficiency of the
elliptic solver.

\subsection{A black hole binary}\label{sec:bbh}

Finally, we solve a classic black hole binary (BBH) initial data problem, which
stands at the beginning of every BBH simulation performed with the \spec{}
code. Again, we solve the full Einstein constraint system in the XCTS formulation,
\cref{eq:xcts}, but now we choose background quantities and boundary conditions
that represent two black holes in orbit. Following the formalism for
\emph{superposed Kerr-Schild} initial data, e.g., laid out in
\ccite{Lovelace2008-sw, Varma2018-fp}, we set the conformal metric and the trace
of the extrinsic curvature to the superpositions
\begin{subequations}\label{eq:bbh_bg}
\begin{align}
  \bar{\gamma}_{ij} &= \delta_{ij} + \sum_{n=1}^2 e^{-r_n^2 / w_n^2} \, (\gamma_{ij}^{(n)} - \delta_{ij})
\intertext{and}
  K &= \sum_{n=1}^2 e^{-r_n^2 / w_n^2} K^{(n)}
  \text{,}
\end{align}
\end{subequations}
where $\gamma_{ij}^{(n)}$ and $K^{(n)}$ are the conformal metric and
extrinsic-curvature trace of two isolated Schwarzschild black holes in
Kerr-Schild coordinates as given in \cref{eq:kerrschild_bg}. They have mass
parameters~$M_n$ and are centered at coordinates~$\bm{C}_n$, with $r_n$ being
the Euclidean coordinate distance from either center. The superpositions are
modulated by two Gaussians with widths $w_n$. The time-derivative quantities
$\bar{u}_{ij}$ and $\partial_t K$ in the XCTS equations~\eqref{eq:xcts} vanish,
as do the matter sources $\rho$, $S$ and~$S^i$.

To handle orbital motion we split the shift in a \emph{background} and an
\emph{excess} contribution~\cite{Pfeiffer2003-mt},
\begin{equation}\label{eq:shift_split}
  \beta^i = \beta^i_\mathrm{background} + \beta^i_\mathrm{excess}
  \text{,}
\end{equation}
and choose the background shift
\begin{equation}\label{eq:bbh_bg_shift}
  \beta^i_\mathrm{background} = (\bm{\Omega}_0 \times \bm{x})^i
  \text{,}
\end{equation}
where $\bm{\Omega}_0$ is the orbital angular velocity. We insert
\cref{eq:shift_split} in the XCTS equations~\eqref{eq:xcts} and henceforth solve
them for $\beta^i_\mathrm{excess}$, instead of $\beta^i$.

\begin{figure}
  \centering
  \subfloat[Black hole binary domain\protect\label{fig:bbh_domain_far}]{
    \includegraphics[width=0.9\columnwidth]{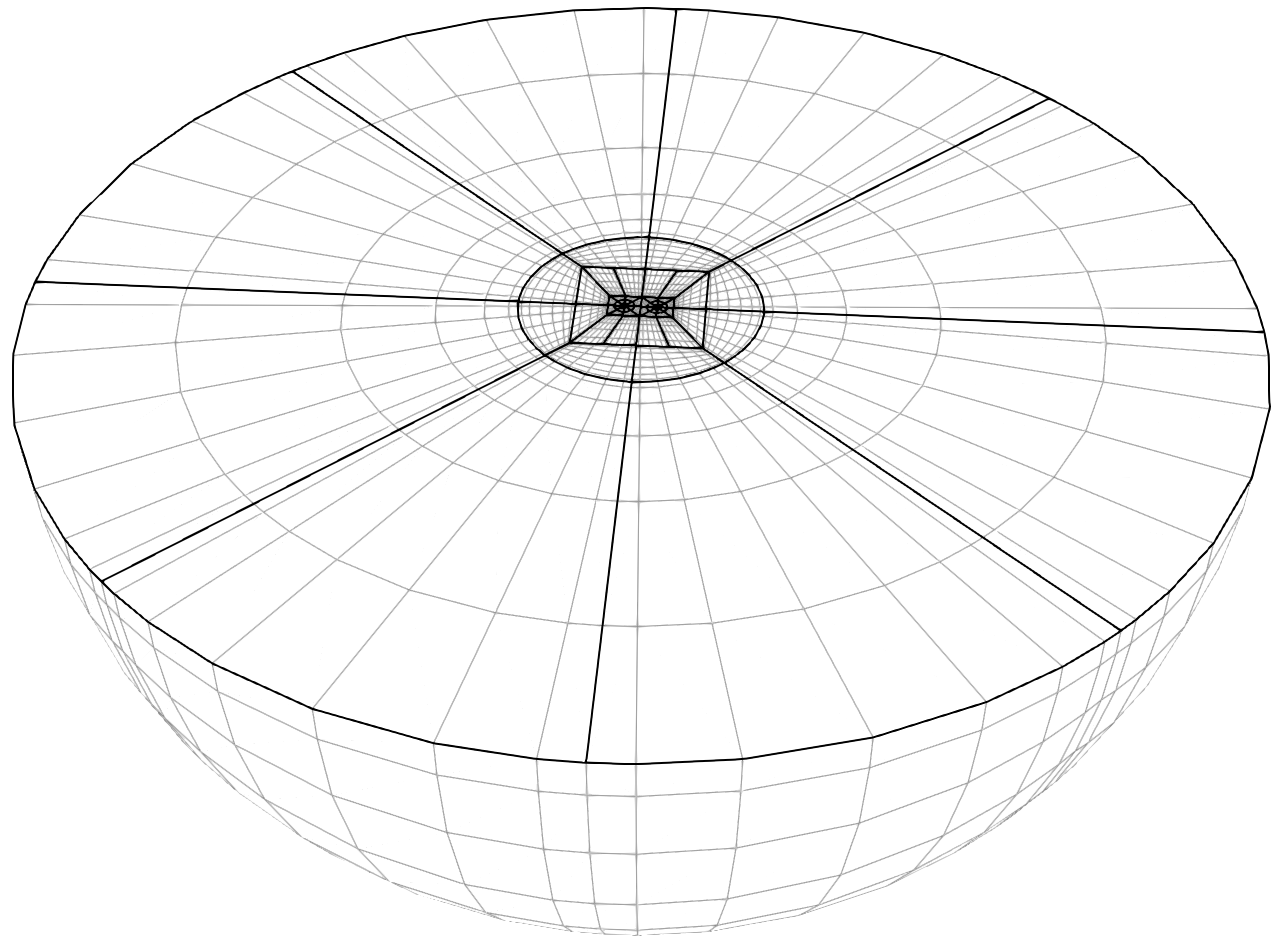}
  }\\
  \subfloat[Close-up\protect\label{fig:bbh_domain_close}]{
    \includegraphics[width=0.9\columnwidth]{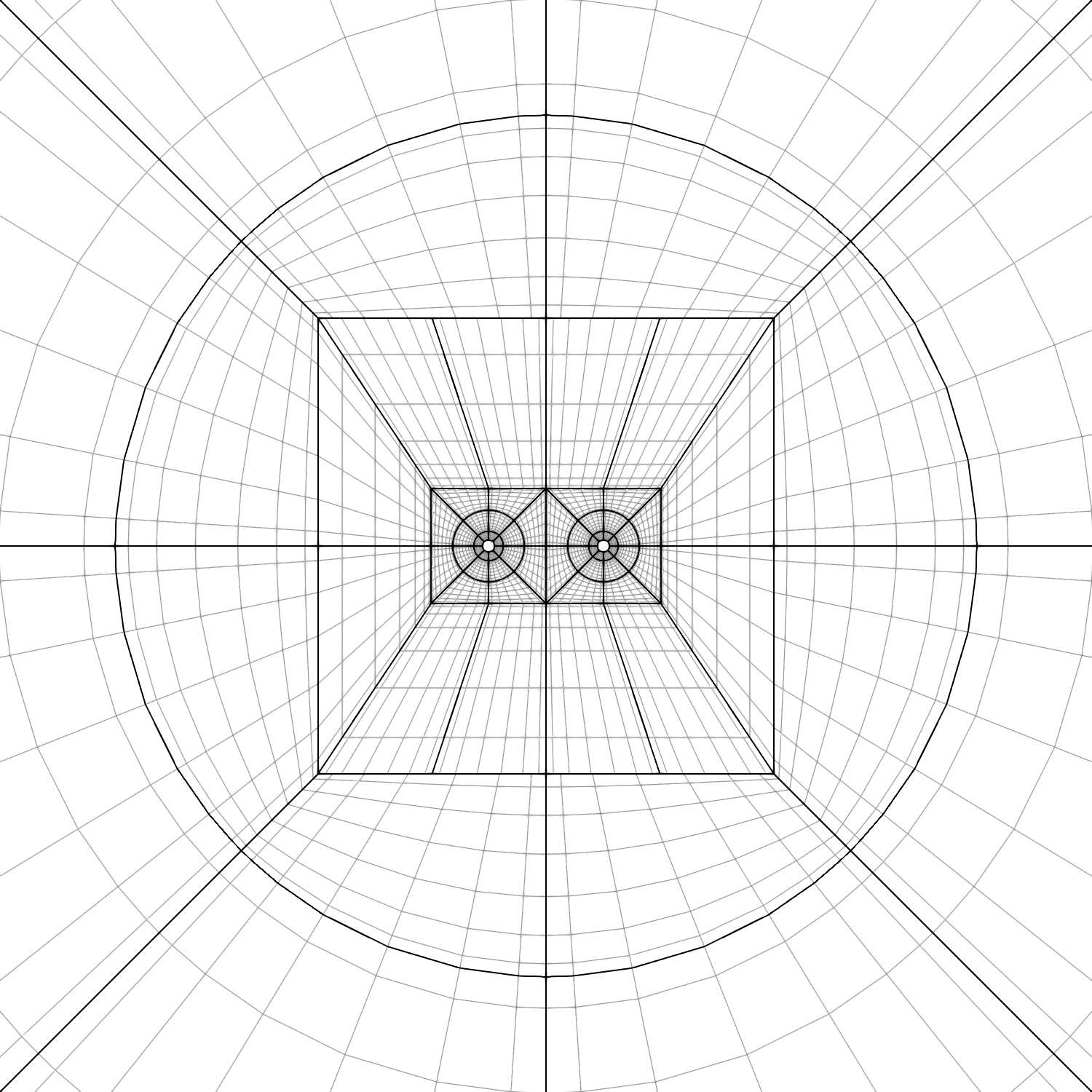}
  }
  \caption{
    \label{fig:bbh_domain}
    A cut through the three-dimensional black hole binary domain used in
    \cref{sec:bbh}. It involves two excised spheres centered at $\bm{C}_n$ along
    the $x$-axis and extends to a spherical outer surface at radius $R$. The
    domain is $h$-refined such that spherical wedges have equal angular size, so
    the cube-to-sphere boundary is nonconforming. All elements in this picture
    have eight angular grid points, and $\{7, 8, 8, 9, 11, 11\}$ radial grid
    points in the layers ordered from outermost to innermost.}
\end{figure}

We solve the XCTS equations on the domain depicted in \cref{fig:bbh_domain}. It
has two excised spheres with radius $2 M_n$ that are centered at $\bm{C}_n$, and
correspond to the two black holes, and an outer spherical boundary at finite
radius~$R$. We impose boundary conditions on these three boundaries as follows.
At the outer spherical boundary of the domain we impose asymptotic flatness,
\begin{equation}\label{eq:flatness_bc}
  \psi = 1 \text{,} \quad \alpha \psi = 1 \text{,} \quad \beta_\mathrm{excess}^i = 0
  \text{.}
\end{equation}
Since the outer boundary is at a finite radius, the solution will only be
approximately asymptotically flat.
On the two excision boundaries we impose (nonspinning) quasiequilibrium
apparent-horizon boundary conditions~\cite{Cook2004-yf}
\begin{subequations}\label{eq:ah_bc}
\begin{align}
  \bar{s}^k\partial_k\psi ={} &-\frac{\psi^3}{8\alpha}\bar{s}_i \bar{s}_j\left((\bar{L}\beta)^{ij} - \bar{u}^{ij}\right) \nonumber \\
  &- \frac{\psi}{4}\bar{m}^{ij}\bar{\nabla}_i \bar{s}_j + \frac{1}{6}K\psi^3 \text{,} \\
  \beta^i ={} &\frac{\alpha}{\psi^2}\bar{s}^i
  \text{,}
\end{align}
\end{subequations}
where $\bar{m}^{ij}=\bar{\gamma}^{ij}-\bar{s}^i \bar{s}^j$. Here, $\bar{s}_i =
-n_i = \psi^{-2} s_i$ is the conformal surface normal to the apparent horizon,
which is opposite the normal to the domain boundary since both are
normalized with the conformal metric. For the lapse
we choose to impose the isolated solution, \cref{eq:kerrschild_lapse}, as
Dirichlet conditions at both excision surfaces. Note that this choice differs
slightly from \ccite{Varma2018-fp}, where the \emph{superposed} isolated
solutions are imposed on the lapse at both excision surfaces. 

\begin{figure*}
  \centering
  \subfloat{
    \includegraphics[width=\columnwidth]{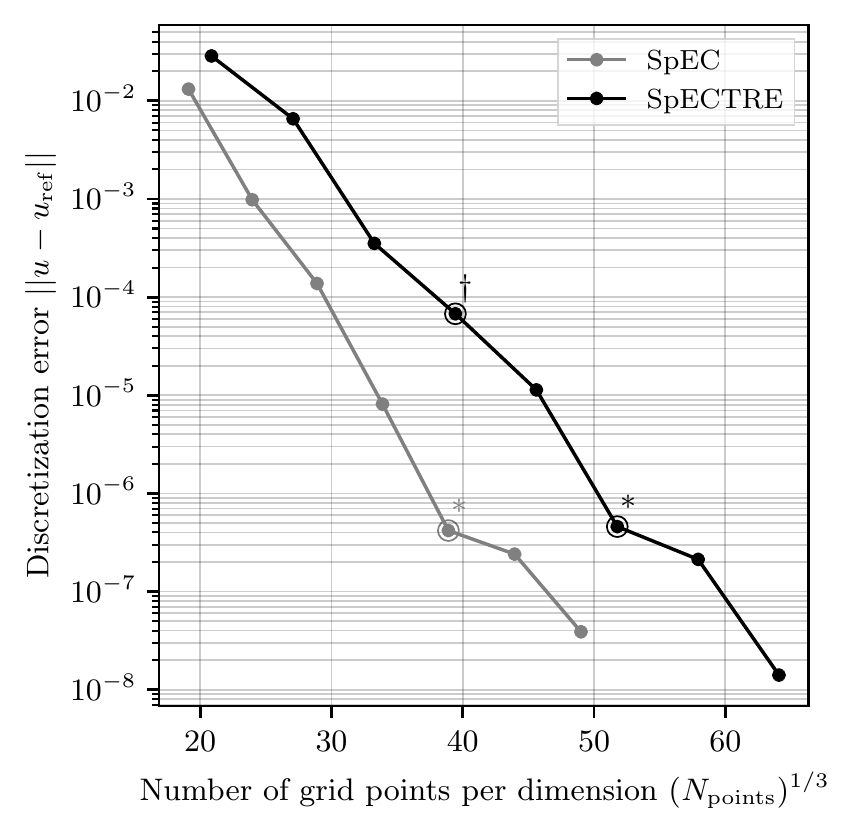}
    }
  \subfloat{
    \includegraphics[width=\columnwidth]{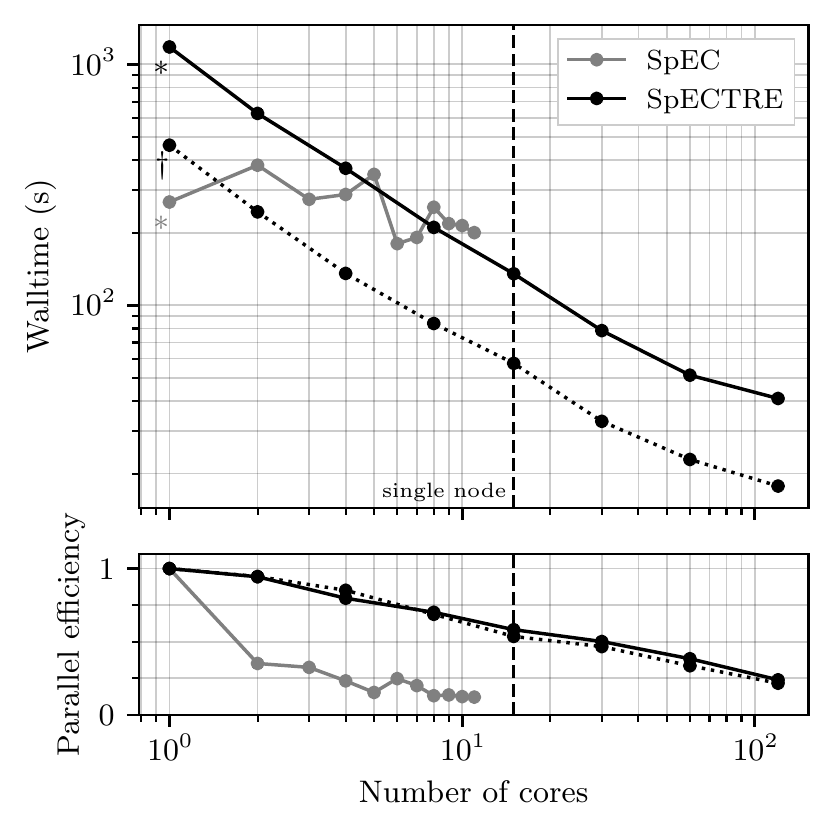}
    }
  \caption{
    \label{fig:bbh_comparison}
    Comparison of the BBH initial data problem (\cref{sec:bbh})
    solved with our new elliptic solver in \spectre{} (black), and with the
    \spec{} elliptic solver (gray). \emph{Left:} Both codes converge
    exponentially with resolution. \spectre{} needs about \SI{30}{\percent} more
    grid points per dimension than \spec{} to reach the same accuracy.
    \emph{Right:} Parallel scaling of both codes. The \spec{} elliptic
    solver scales to at most eleven cores and reaches a speedup of at most a
    factor of two compared to the single-core runtime. Our new elliptic solver
    in \spectre{} is faster than \spec{} on eight cores, and scales the
    problem reliably to 120 cores, at which point it is seven times faster than
    \spec{}'s single-core runtime. The dotted line corresponds to a
    configuration with the same number of grid points as \spec{} (but
    lower accuracy), which is faster than \spec{} on only two cores.}
\end{figure*}

To assess the accuracy and parallel performance of the elliptic solver for the
BBH initial data problem we solve the same scenario with the
\spec{}~\cite{Pfeiffer2003-mt,spec} code. In \spec{} we successively increment the resolution
from \texttt{Lev0} to \texttt{Lev6}, which correspond to domain configurations
determined with an adaptive mesh-refinement (AMR) algorithm. In \spectre{} we
simply increment the number of grid points in all dimensions of all elements by one
from each resolution to the next, based on the domain depicted in
\cref{fig:bbh_domain}. To compare the solution between the two codes, we
interpolate all five fields $u_A = \{\psi, \alpha\psi,
\beta_\mathrm{excess}^i\}$ to a set of sample points $\bm{x}_m$. We do the same
for a very high-resolution run with \spectre{} that we use as reference,
$u_{A,\mathrm{ref}}$, for which we have split every element in the domain in two
along all three dimensions. Then, we compute the discretization error for all
\spec{} and \texttt{SpECTRE} solutions as an $L_2$-norm of the difference
to the high-resolution reference run over all fields and sample points,
\begin{equation}
  \lVert u - u_\mathrm{ref} \rVert \defeq \left(
    \sum_{A,m} \left(u_A(\bm{x}_m) - u_{A,\mathrm{ref}}(\bm{x}_m)\right)^2
  \right)^{1/2}
  \mkern-22mu \text{.}
\end{equation}
We have chosen $M_n=0.4229$, $\bm{C}_n=(\pm 8, 0, 0)$, $\Omega_0=0.0144$,
$w_n=4.8$, $R=300$, and sample points along the $x$-axis at $x_1=8.846$
(near horizon), $x_2=0$ (origin) and $x_3=100$ (far field) here. This
configuration coincides with our convergence study in \ccite{dgscheme}, where we
list the reference values $u_{A,\mathrm{ref}}(\bm{x}_m)$ at the interpolation
points explicitly.

\Cref{fig:bbh_comparison} compares the performance of the BBH
initial data problem with the \spec{} code. Both \spec{} and \spectre{} converge
exponentially with resolution, since \spec{} employs a spectral scheme and
\spectre{} a DG scheme under $p$~refinement. \spectre{} currently needs about
\SI{30}{\percent} more grid points per dimension to achieve the same accuracy as
\spec{}. To an extent this is to be expected, since we split the domain into
more elements than \spec{} does and hence have more shared element boundaries
with duplicate points. In particular, \spec{} employs shells with spherical
basis functions that avoid duplicate points in angular directions altogether.
While the \spec{} elliptic solver always decomposes the domain into eleven
subdomains, each with up to 34 grid points per dimension in our test, our domain
in \spectre{} has 232 elements with up to 13 grid points per dimension. However,
neither is our BBH domain in \spectre{} optimized as well as
\spec{}'s yet, nor have we refined it with an AMR algorithm. We are planning to
do both in future work. Furthermore, \spec{}'s initial data domain involves
overlapping patches to enable matching conditions for its spectral scheme, which
our DG scheme in \spectre{} does not need. Therefore, we expect to achieve
domain configurations that come closer to \spec{} in their number of grid points
with future optimizations.

The right panel in \cref{fig:bbh_comparison} demonstrates the superior parallel
performance that our new elliptic solver achieves over \spec{}'s. We choose the
runs marked with~$^*$ in the left panel because they solve the BBH
problem to comparable accuracy. We scale these runs to an increasing number of
cores and measure the wall time for the elliptic solves to complete. Since the
\spec{} initial data domain is composed of eleven subdomains, it can parallelize
to at most eleven cores. The runtime decreases by a factor of about 1.5 to 2 when
the solve is distributed to multiple cores, but shows little reliable scaling.
Some \spec{} configurations at higher resolutions have shown slightly better
parallel performance, but none that exceeded a factor of about
two in speedup compared to the single-core runtime. Our new elliptic solver in
\spectre{}, on the other hand, scales reliably to 120 cores, at which point each
core holds no more than two elements. On a single core it needs
\SI{1176}{\second} where \spec{}, with fewer grid points, needs only
\SI{268}{\second}, but it overtakes \spec{} on eight cores and completes in only
\SI{37}{\second} on 120 cores. For reference we have also included a scaling
test with \spectre{} that uses the same number of grid points as \spec{} but
does not yet reach the same accuracy (marked with~$^\dagger$). It overtakes
\spec{} on two cores and completes in only \SI{14}{\second} on 120 cores. The
configuration represents a potential improvement in the domain decomposition
with future optimizations. We find the parallel efficiency for the BBH
configurations behaves similarly to the single black hole configurations
we investigated in \cref{sec:kerrschild}.

\section{Conclusion and future work}\label{sec:conclusion}

We have presented a new solver for elliptic partial differential equations that
is designed to parallelize on computing clusters. It is based on discontinuous
Galerkin (DG) methods and task-based parallel iterative algorithms. We have
shown that our solver is capable of parallelizing elliptic problems with
\num{\sim 200} million degrees of freedom to at least a few thousand cores. It
solves a classic black hole binary (BBH) initial data problem faster than the veteran
code \spec{}~\cite{spec} on only eight cores, and in a fraction of the time when
distributed to more cores on a computing cluster. The elliptic solver is
implemented in the open-source \spectre{}~\cite{spectre} numerical relativity code, and the
results in this article are reproducible with the supplemental input-file
configurations.\footnote{\url{https://arxiv.org/src/2111.06767/anc}}

So far we can solve Poisson, elasticity, puncture and XCTS problems, including
BBH initial data in the superposed Kerr-Schild formalism with
unequal masses, spins and negative-expansion boundary
conditions~\cite{Varma2018-fp} (in this article we only explored an equal-mass
and nonspinning BBH). In the short term we are planning to add
the capability to solve for binary neutron star (BNS) and black hole--neutron star (BHNS)
initial data, which involve the XCTS equations coupled to the equations of
hydrostatic equilibrium.

A notable strength of our new elliptic solver is the multigrid-Schwarz
preconditioner, which achieves iteration counts independent of the number of
elements in the computational domain. Therefore, we expect our solver to scale
to problems that benefit from $h$~refinement, e.g., to resolve different length
scales or to adapt the domain to features in the solution. Such problems include
initial data involving neutron stars with steep gradients near the surface,
equations of state with phase transitions, or simulating thermal noise in thin
mirror coatings for gravitational-wave detectors~\cite{Lovelace2017xyf,
thermalnoise}.

Our solver splits the computational domain into more elements than the spectral
code \spec{} to achieve superior parallelization properties. However, the larger
number of elements with shared boundaries also means that we need more grid
points than \spec{} to reach the same accuracy for a BBH
initial data problem. Variations of the DG scheme, such as a hybridizable DG
method, can provide a possible resolution to this
effect~\cite{Cockburn2009UnifiedHO, Giacomini2020HDGlabAO,
Muralikrishnan2020AMA}. Even without changing the DG scheme, we expect that
optimizations of our binary compact object domain can significantly reduce the
number of grid points required to reach a certain accuracy. Possible domain
optimizations include combining the enveloping cube and the cube-to-sphere
transition into a single layer of blocks, equalizing the angular size of the
enveloping wedges in a manner similar to \ccite{Rashti:2021ihv}, or more drastic
changes that involve cylindrical or bipolar coordinate maps, such as the domain
presented in \ccite{Buchman2012-ud}. To retain the effectiveness of the
multigrid solver it is important to keep the number of blocks in the
domain to a minimum under these optimizations. We have shown that our new
elliptic solver reaches comparative single-core performance to \spec{} when
using the same number of grid points, with the added benefit of parallel
scaling. Since every contemporary computer has multiple cores, we prioritize
parallelization over single-core performance.

To put the grid points of the computational domain to most effective use,
adaptive mesh-refinement (AMR) techniques will be essential. All components of
the elliptic solver, including the DG discretization, the multigrid algorithm,
and the Schwarz subdomains, already support $hp$-refined domains. The refinement
can be anisotropic, meaning elements can be split or differ in their polynomial
degree along each dimension independently. A major subject of future work will
be the development of an AMR scheme that adjusts the refinement during the
elliptic solve automatically based on a local error estimate, distributing
resolution to regions in the domain where it is most needed.

Along with AMR, we expect load balancing to become increasingly important. We
currently approximately load balance the elliptic solver at the beginning of the
program based on the number of grid points in each element. \texttt{Charm++},
and hence \spectre{}, also support dynamic load-balancing operations that migrate
elements between cores periodically, or at specific points in the algorithm.
\texttt{Charm++} provides a variety of load-balancing algorithms that may take
metrics such as runtime measurements, communication cost and the network
topology into account. When the computational load on elements changes due to
$p$-AMR, or when elements get created and destroyed due to $h$-AMR, we intend to
invoke load balancing to improve the parallel performance of the elliptic
solves.

The elliptic solver algorithms can be improved in many ways. The multigrid
solver may benefit from an additive variant of the algorithm, which smooths
every level independently and combines the solutions~\cite{AlOnazi:2017}. An
additive multigrid algorithm has better parallelization properties than the
multiplicative algorithm that we employ in this article, since coarse grids do
not need to wait for fine grids to send data before the coarse-grid smoothing
can proceed. However, the additive multigrid algorithm typically requires more
iterations to converge than the multiplicative. Furthermore, multigrid patterns
other than the standard \vcycle{} may accelerate convergence, such as a W\=/cycle or
F\=/cycle pattern~\cite{Briggs2000jp}.

Schwarz solvers also come in many variations, e.g., involving face-centered
subdomains, that we have not explored in this article. Our element-centered
subdomains that eliminate corner and edge neighbors have served well for our
DG-discretized problems so far, and we have focused on accelerating the
subdomain solves with suitable preconditioners. Faster explicit-construction and
approximate-inversion techniques for the subproblems in the
Laplacian-approximation preconditioner have the greatest potential to speed up
the elliptic solver. Possibilities include constructing matrix representations
analytically, either from the DG scheme or from an approximate finite-difference
scheme, and faster methods to solve the subproblems than the incomplete LU
technique we currently employ.

A possible avenue for a more drastic improvement of the elliptic solver
algorithm is to replace the multigrid-Schwarz preconditioner, or parts of it,
altogether. For example, recent developments in the field of physics-informed
neural networks (PINNs) suggest that hybrid strategies, combining a traditional
linear solver with a PINN, can be very effective \cite{Markidis2021,
guidetti2021dnnsolve}. Hence, an intriguing prospect for accelerating elliptic
solves in numerical relativity is to combine our Newton-Krylov algorithm with a
PINN preconditioner, use the PINN as a smoother on multigrids, or use it to
precondition Schwarz subdomain solves.

Looking ahead, fast, scalable and highly-parallel elliptic solves in numerical
relativity not only have the potential to accelerate initial data construction
to seed high-resolution simulations of general-relativistic scenarios, and at
extreme physical parameters, but they may also support evolutions.
For example, some gauge constraints can be formulated as elliptic equations, and
solving them alongside an evolution can allow the choice of beneficial
coordinates, such as maximal slicing~\cite{BaumgarteShapiro}. The
apparent-horizon condition is also an elliptic equation, though current NR codes
typically find apparent horizons with a parabolic relaxation method~\cite{Gundlach1997us}. Some NR
codes employ a constrained-evolution scheme, which evolves the system in time
through a series of elliptic solves, or employ implicit-explicit (IMEX)
evolution schemes~\cite{Lau2011we}. Lastly, Einstein-Vlasov systems for collisionless matter
involve elliptic equations, as do simulations that involve solving a Poisson
equation alongside an evolution, such as simulations of self-gravitating
protoplanetary disks~\cite{Deng2020,Hopkins:2014qka,Enzo2014}. Currently, elliptic solvers are rarely applied to solve
any of these problems alongside an evolution because they are too costly. Fast
elliptic solves have the potential to enable these applications.

\begin{acknowledgments}
The authors thank Tim Dietrich, Francois Foucart, and Hannes R{\"u}ter for
helpful discussions. N.~V.\ also thanks the Cornell Center for Astrophysics and
Planetary Science and TAPIR at Caltech for the hospitality and financial support
during research stays.
Computations were performed with the \spectre{} code~\cite{spectre} on the
\texttt{Minerva} cluster at the Max Planck Institute for Gravitational Physics.
\texttt{Charm++}~\cite{charmpp} was developed by the Parallel Programming
Laboratory in the Department of Computer Science at the University of Illinois
at Urbana-Champaign.
The figures in this article were produced with \texttt{dgpy}~\cite{dgpy},
\texttt{matplotlib}~\cite{matplotlib1,matplotlib2}, \texttt{TikZ}~\cite{tikz}
and \texttt{ParaView}~\cite{paraview}.
This work was supported in part by the Sherman Fairchild Foundation and by NSF
Grants No.\ PHY-2011961, No.\ PHY-2011968, and No.\ OAC-1931266 at Caltech, and NSF
Grants No.\ PHY-1912081 and No.\ OAC-1931280 at Cornell. G.~L.\ is pleased to
acknowledge support from the NSF through Grants No.\ PHY-1654359 and No.\
AST-1559694 and from Nicholas and Lee Begovich and the Dan Black Family Trust.
\end{acknowledgments}

\bibliography{References}

\end{document}